\pdfoutput=1
\newcommand*{\ATLASLATEXPATH}{latex/}
\documentclass[UKenglish,cernpreprint,texlive=2016,atlasstyle=true]{\ATLASLATEXPATH atlasdoc}

\usepackage[utf8]{inputenc}
\DeclareUnicodeCharacter{00A0}{ }
\DeclareUnicodeCharacter{00A0}{~}
\usepackage{\ATLASLATEXPATH atlaspackage}
\usepackage{\ATLASLATEXPATH atlasbiblatex}
\usepackage{\ATLASLATEXPATH atlascontribute}

\usepackage{\ATLASLATEXPATH atlasphysics}
\usepackage{multirow}
\usepackage{caption}
\graphicspath{{logos/}{figures/}}

\newcommand{\sNN}{\ensuremath{\sqrt{s_{NN}}}\xspace}
\newcommand{\vn}{\ensuremath{v_{n}}\xspace}

\newcommand{\vtwo}{\ensuremath{v_{2}}\xspace}
\newcommand{\vthree}{\ensuremath{v_{3}}\xspace}
\newcommand{\vfour}{\ensuremath{v_{4}}\xspace}
\newcommand{\vfive}{\ensuremath{v_{5}}\xspace}
\newcommand{\vsix}{\ensuremath{v_{6}}\xspace}
\newcommand{\vseven}{\ensuremath{v_{7}}\xspace}
\newcommand{\fcalsum}{\ensuremath{\Sigma E_\mathrm{T}^\mathrm{FCal}}\xspace}

\newcommand{\mub}{\ensuremath{\mu\mathrm{b}}\xspace}
\newcommand{\cent}[1]{#1\%\xspace}

\newlength{\wone}
\newlength{\wtwo}
\newlength{\wsmalltwo}
\newlength{\wthree}
\newlength{\wfour}
\newlength{\wfive}
\newcommand{\includetwo}[2]{
  \includegraphics[width=\wtwo]{#1}%
  \if\relax\detokenize{#2}\relax
  \else
  \includegraphics[width=\wtwo]{#2}
  \fi  
  \\}

\newcommand{\includesmalltwo}[2]{
  \includegraphics[scale=0.27]{#1}%
  \if\relax\detokenize{#2}\relax
  \else
   \hspace{15pt} 
  \includegraphics[scale=0.27]{#2}
  \fi  
  \\}

\newcommand{\includethree}[3]{
  \includegraphics[width=\wthree]{#1}%
  \if\relax\detokenize{#2}\relax
  \else
  \includegraphics[width=\wthree]{#2}%
  \fi
  \if\relax\detokenize{#3}\relax
  \else
  \includegraphics[width=\wthree]{#3}
  \fi
  \\}

\newcommand{\includefour}[4]{
  \includegraphics[width=\wfour]{#1}%
  \includegraphics[width=\wfour]{#2}%
  \if\relax\detokenize{#3}\relax
  \else
  \includegraphics[width=\wfour]{#3}%
  \fi
  \if\relax\detokenize{#4}\relax
  \else
  \includegraphics[width=\wfour]{#4}
  \fi
  \\}

\newcommand{\PbPb}{\mbox{Pb+Pb}\xspace}

\newcommand{\Ncoll}{\mbox{$N_{\mathrm{coll}}$}}

\newcommand{\Njet}{\mbox{$N_{\mathrm{jet}}$}}

\newcommand{\Npart}{\mbox{$N_{\mathrm{part}}$}}

\newcommand{\sqrtsnn}{\mbox{$\sqrt{s_{_\text{NN}}}$}}

\newcommand{\TAA}{\mbox{$T_{\mathrm{AA}}$}}

\newcommand{\pTjet}{\mbox{$p_{{\mathrm{T}}}$}}

\newcommand{\pTtrk}{\mbox{$p_{{\mathrm{T}}}^{\mathrm{trk}}$}}

\newcommand{\pTtrue}{\mbox{$p_{\mathrm{T}}^{\mathrm{truth}}$}}

\newcommand{\ETjet}{\mbox{$\ET^{\mathrm{jet}}$}}

\newcommand{\ptjet}{\mbox{$p_{\mathrm{T}}^{\mathrm{jet}}$}}

\newcommand{\dphi}{\ensuremath{\Delta\phi}\xspace}

\newcommand{\vnn}{\ensuremath{v_{n,n}}\xspace}
\newcommand{\voo}{\ensuremath{v_{1,1}}\xspace}
\newcommand{\vtt}{\ensuremath{v_{2,2}}\xspace}
\newcommand{\vthth}{\ensuremath{v_{3,3}}\xspace}
\newcommand{\vff}{\ensuremath{v_{4,4}}\xspace}
\newcommand{\vfifi}{\ensuremath{v_{5,5}}\xspace}
\newcommand{\vsisi}{\ensuremath{v_{6,6}}\xspace}

\newcommand{\ptab}{\ensuremath{p_{\mathrm{T}}^{a,b}}\xspace}
\newcommand{\pta}{\ensuremath{p_{\mathrm{T}}^{a}}\xspace}
\newcommand{\ptb}{\ensuremath{p_{\mathrm{T}}^{b}}\xspace}
\newcommand{\phia}{\ensuremath{\phi^a}\xspace}
\newcommand{\phib}{\ensuremath{\phi^b}\xspace}
\newcommand{\etab}{\ensuremath{\eta^b}\xspace}
\newcommand{\deta}{\ensuremath{\Delta\eta}\xspace}
\newcommand{\etaaa}{\ensuremath{\eta^a}\xspace}
\newcommand{\dphia}{\ensuremath{\Delta\phi}\xspace}

\newcommand{\ctwo}{\ensuremath{C(\Delta\eta,\Delta\phi)}\xspace}
\newcommand{\ctwophi}{\ensuremath{C(\Delta\phi)}\xspace}

\newcommand{\papertype}{paper\xspace}

\newcommand{\tpc}{2PC\xspace}

\usepackage{\ATLASLATEXPATH atlasphysics}
\AtlasTitle{Measurement of the azimuthal anisotropy of charged particles produced in $\sqrtsnn$ = 5.02~TeV Pb+Pb collisions with the ATLAS detector}

\author{The ATLAS Collaboration}

\AtlasRefCode{HION-2016-06}

\AtlasNote{HION-2016-06}

\PreprintIdNumber{CERN-EP-2018-194}

\AtlasDate{\today}

\AtlasJournal{EPJC}
\AtlasJournalRef{\EPJC 78 (2018) 997}
\AtlasDOI{10.1140/epjc/s10052-018-6468-7}

\AtlasAbstract{%
Measurements of the azimuthal anisotropy in lead--lead collisions at \sqrtsnn\ = 5.02~TeV are presented using a data sample corresponding to 0.49 $\mathrm{nb}^{-1}$ integrated luminosity collected by the ATLAS experiment at the LHC in 2015. The recorded minimum-bias sample 
is enhanced by triggers for ``ultra-central'' collisions, providing an opportunity to perform detailed study of flow harmonics in the regime where the initial state is dominated by fluctuations.\ The anisotropy of the charged-particle azimuthal angle distributions is characterized by the Fourier coefficients, $v_{2}$--$v_{7}$, which are measured using the two-particle correlation, scalar-product and event-plane  methods.
The goal of the paper is to provide measurements of the differential as well as integrated flow harmonics \vn over wide ranges of the transverse momentum, 0.5~$<\pT<$~60~GeV, the pseudorapidity, $|\eta|<$~2.5, and the collision centrality 0--80\%. Results from different methods are compared and discussed in the context of previous and recent measurements in Pb+Pb collisions at \sqrtsnn\ =~2.76~\TeV\ and 5.02~\TeV.
In particular, the shape of the \pT dependence of elliptic or triangular flow harmonics is observed to be very similar at different centralities after scaling the \vn and \pT values by constant factors over the centrality interval 0--60\% and the \pT range 0.5~$< \pT <$~5~GeV.
}

\AtlasCoverSupportingNote{}

\hypersetup{pdftitle={ATLAS draft},pdfauthor={The ATLAS Collaboration}}

\begin{document}
\maketitle

\section{Introduction}
\label{sec:intro}
One of the primary goals of ultra-relativistic heavy-ion collisions is the study of the hot and dense medium formed there, usually referred to as  the  quark-gluon plasma (QGP)~\cite{Back:2004je,Adams:2005dq,Arsene:2004fa,Adcox:2004mh, Schukraft:2017nbn}.\
The existence of the QGP phase of nuclear matter has been confirmed by a wealth of 
   experimental data~\cite{Schukraft:2017nbn,FOKA2016154}.\ 
In particular, properties related to the collective expansion of 
   the QGP (e.g.\ the equation of state~\cite{PhysRevLett.114.202301} 
   and shear viscosity~\cite{PhysRevLett.115.132301}) are 
   inferred from measurements of azimuthal anisotropies of produced 
   particles.\
It is now understood that the azimuthal anisotropy results from large 
   initial pressure gradients in the hot, dense matter created in 
   the collisions~\cite{PhysRevD.46.229,Voloshin:1994mz}.\
   These pressure gradients transform the initial spatial anisotropies of 
   nuclear collisions into momentum anisotropies of the final-state 
   particle production, which are experimentally characterized by 
   Fourier (flow) harmonics of the azimuthal angle distributions of produced 
   particles.\
The discovery of large flow harmonics at RHIC, and more recently at 
   much higher collision energy at the 
   LHC~\cite{PhysRevLett.105.252302,HION-2011-05,HION-2011-01,CMS-HIN-10-002}, 
   has significantly deepened the understanding of the QGP, 
   as explored theoretically by the QCD 
   lattice~\cite{lattice}.\
In particular, the recent measurements of azimuthal anisotropy help 
   to constrain the commonly used modelling of the dynamics of
   heavy-ion collisions based on relativistic viscous hydrodynamics.\ 
Typically, in the hydrodynamic models, a strongly interacting quark--gluon 
   medium is formed shortly after the collision and its evolution is well 
   described by relativistic fluid dynamics~\cite{PhysRevLett.115.132301}.\
Detailed investigations, based on hydrodynamics, have shown that the 
   produced medium has properties similar to those of an almost ideal fluid 
   characterized by a very low ratio of viscosity to entropy density, $\eta/s$.\   
Precise azimuthal anisotropy measurements over a wide range in kinematic 
  variables and centrality are key elements to improving our understanding of 
  the strongly coupled QGP because of their unique sensitivity to $\eta/s$.

The azimuthal angular distribution of single produced particles can be 
   expanded in a Fourier series~\mbox{\cite{Poskanzer:1998yz,Voloshin:2008dg}}:
\begin{equation}
\frac{\text{d}N}{\text{d}\phi} = \frac{N_{0}}{2\pi}\left( 1+ \sum_{n=1} 2\vn\cos\left[n\left(\phi - \Phi_n\right)\right] \right),
\label{eq:single}
\end{equation}
where $N_0$ is the total particle yield,
   $\phi$ is the azimuthal angle of the produced particles 
   and the $\vn$ and $\Phi_n$ are, respectively, the magnitude of the 
   $n^{\mathrm{th}}$-order azimuthal anisotropy and the orientation of 
   the $n^{\mathrm{th}}$-order symmetry plane.\
The \vn coefficients -- also called \textit{flow harmonics} -- are typically 
   measured as a function of particle
   pseudorapidity\footnote{ATLAS
    uses a right-handed coordinate system with its origin at the
    nominal interaction point (IP) in the centre of the detector
    and the $z$-axis along the beam pipe.\
     The $x$-axis points from
    the IP to the centre of the LHC ring, and the $y$-axis points upward.
    Cylindrical coordinates $(r,\phi)$ are used in the transverse plane,
    $\phi$ being the azimuthal angle around the $z$-axis.\
    The pseudorapidity is defined in terms of the polar angle $\theta$
    as $\eta=-\ln\tan(\theta/2)$.} ($\eta$), transverse momentum (\pT), and 
   the degree of overlap between the colliding nuclei (centrality).\
Event-by-event fluctuations in the number and position of the interacting 
nucleons give rise to anisotropic flow fluctuations~\cite{fluctCGC}.

The first harmonic, $v_{1}$, is known as \textit{directed flow} and 
   refers to the sideward motion of participants in ultra-relativistic
   nuclear collisions, and it carries information from the 
   early stage of the collision.\
The most extensive studies are related to the second flow harmonic, 
   \vtwo, also known as \textit{elliptic flow}.\ 
Elliptic flow  is sensitive to the initial spatial asymmetry of the 
   almond-shaped overlapping zone of the colliding nuclei.\
The higher-order coefficients, $n>2$, are also important
  due to their sensitivity to initial-state geometric fluctuations 
  and viscous effects~\cite{Poskanzer:1998yz,Voloshin:2008dg,fluctCGC}.

During the first operational period at the LHC (Run\,1) lead ions 
   were collided at energy per colliding nucleon--nucleon pair \mbox{\sqrtsnn~=~2.76~\TeV}, 
   which is about 13 times larger than the highest collision energy 
   attained at RHIC in Au+Au collisions.\
ATLAS and other LHC experiments collected large samples of heavy-ion 
   data enabling extensive studies of the elliptic flow and
   higher-order Fourier coefficients.\ 
ATLAS measurements of flow harmonics were performed in broad 
   regions of transverse momentum, pseudorapidity and event centrality, 
   using the standard event-plane (EP) method~\cite{HION-2011-05}, 
   two-particle correlations (2PC)~\cite{HION-2011-01} and 
   multi-particle cumulants~\cite{HION-2012-04}.\
Significant (non-zero) flow harmonics up to \vsix were measured
   in Pb+Pb collisions at \sqrtsnn~=~2.76~\TeV, which provide 
   important constraints on the bulk and shear viscosity of the 
   QGP medium~\cite{Charlie:2015}.\
Additionally, by comparing RHIC (STAR~\cite{Adams:2004bi} and 
   PHENIX~\cite{Adare:2010sp}) and LHC (ATLAS~\cite{HION-2011-05}, 
   ALICE~\cite{Aamodt:2010pa} and CMS~\cite{Chatrchyan:2012ta}) 
   results, it was found that  for similar centrality of Au+Au and Pb+Pb interactions, \vn 
   as a function of \pT is  approximately independent of collision energy.\ 
There is an initial rise of \vn with \pT up to about 3~\GeV\ 
  and then a drop-off at higher values of \pT, 
  and only weak dependence for \pT $>$ 8--9~\GeV.\
As a function of centrality, there is similarly little variation with
  collision energy.\ 
The second harmonic, \vtwo, exhibits the most pronounced centrality variation, 
  rising to a maximum for mid-central collisions, and then falling off for 
  the most central collisions, reflecting variations in 
  the shape of the initial collision geometry.\ 
The harmonic, \vthree, referred as triangular flow, which has a value similar to \vtwo in central 
  collisions, shows a weaker 
  dependence on centrality, as do the higher-order harmonics.

At the start of the second operational period of the LHC (Run\,2), 
   in November and December of 2015, lead--lead collisions with 
   higher collision energy per nucleon pair of $\sqrtsnn= 5.02$~\TeV\ were
   collected by the LHC experiments.\ 
The goal of this \papertype is to present and discuss the first ATLAS 
   measurements of \vn harmonics at this energy, using the two-particle 
   correlation~\cite{Voloshin:2008dg}, scalar-product (SP)~\cite{Adler:2002pu} 
   and event-plane~\mbox{\cite{Poskanzer:1998yz,Voloshin:2008dg}} methods.
Comparing the 2PC and SP results can quantify the extent to which the two-particle correlations
 factorize into the product of the flow harmonics corresponding to 
   single-particle angular distributions~\cite{HION-2016-04,Jia:2014vja}.\ 
While the SP and EP methods are expected to yield similar values of the \vn, 
  small variations due to their different sensitivity to 
   initial-state geometric fluctuations can nevertheless occur~\cite{Luzum:2012da}. 
To study the energy dependence, the 2PC and EP flow harmonics are 
   compared with previous ATLAS measurements in 2.76~\TeV\ Pb+Pb 
   collisions~\cite{HION-2011-05,HION-2011-01}.\
The results presented in this paper, together with the results on 
   azimuthal anisotropy from other LHC experiments~\cite{Adam:2016izf,Sirunyan:2017pan}, 
   provide further opportunity to study the properties of the QGP, 
   constrain hydrodynamic models, study transport coefficients and extract 
   the temperature dependence of transport coefficients, including $\eta/s$. 

The organization of this \papertype is as follows.\
Section\,\ref{sec:setup} gives a brief overview of the ATLAS detector 
   and the subsystems that are used in this analysis.\
Section\,\ref{sec:data} describes the 
   datasets, triggers and the offline selection criteria used to 
   select events and charged-particle tracks.\
Section\,\ref{sec:analysis} gives details of the 
   scalar-product, event-plane and two-particle correlation methods, which are used 
   to measure the \vn.\
Section\,\ref{sec:systematics} describes the systematic uncertainties 
   associated with the measured \vn.\
Section\,\ref{sec:results} presents the main results of the analysis, 
   which are the \pT, $\eta$ and centrality dependence of the \vn and 
   comparisons of results from the different methods.\
Section\,\ref{sec:summary} gives a summary of the main results 
   and observations.

\section{Experimental set-up}
\label{sec:setup}
The measurements were performed
  using the ATLAS detector ~\cite{PERF-2007-01} at the LHC. 
  The principal components used in this analysis are the inner detector~(ID), minimum-bias trigger
  scintillators~(MBTS), calorimeter, zero-degree
  calorimeters~(ZDC), and the trigger and data acquisition
  systems.\
The ID detects charged particles within the pseudorapidity range 
  $|\eta|$~<~2.5 using a combination of silicon pixel detectors, including
  the ``insertable B-layer''~\cite{Capeans:1291633,ATLAS:1451888}
  that was installed between Run\,1 and Run\,2, silicon microstrip
  detectors~(SCT), and  a straw-tube transition radiation tracker (TRT),
  all immersed in a 2\,T axial magnetic field~\cite{IDET-2010-01}.\
The MBTS system detects charged particles over $2.07 < |\eta| < 3.86$
   using two scintillator-based hodoscopes on each side of the detector, positioned at 
   $z=\pm3.6~{\mathrm m}$.\
These hodoscopes were rebuilt between Run\,1 and Run\,2.\
The ATLAS calorimeter system consists of a liquid argon (LAr) electromagnetic (EM)
   calorimeter covering $|\eta|<3.2$, a steel--scintillator sampling hadronic
   calorimeter covering $|\eta|<1.7$, a~LAr hadronic calorimeter covering
   $1.5<|\eta|<3.2$, and two LAr electromagnetic and hadronic forward calorimeters
   (FCal) covering $3.2<|\eta|<4.9$.\
The ZDC, situated at approximately $\pm140$~m from the nominal IP, detect neutral
  particles, mostly neutrons and photons, with $|\eta|$~>~8.3.
The ZDC use tungsten plates as absorbers, and quartz rods sandwiched
  between the tungsten plates as the active medium.\
The ATLAS trigger system~\cite{PERF-2011-02} consists of a first-level (L1)
   trigger implemented using a combination of dedicated electronics and
   programmable logic, and a software-based high-level trigger.

\section{Event and track selection}
\label{sec:data}
The Pb+Pb dataset used in this \papertype\ corresponds to an integrated 
  luminosity of 0.49~$\mathrm{nb}^{-1}$.\
Minimum-bias events were selected by two mutually exclusive triggers :
\begin{itemize}
\item Events with smaller impact parameter (semi-central and central collisions)
        were selected by a trigger that required the total transverse energy ($\ET$)
        deposited in the calorimeters at L1 to be above 50~\GeV.
\item Collisions with large impact parameter (peripheral events)
        were selected by a trigger that required the total transverse energy 
        at L1 to be less than 50~\GeV,  one neutron 
        on either side in the ZDC ($|\eta| > 8.3$), and at least one reconstructed 
        track in the ID.
\end{itemize}
The minimum-bias triggers sampled a total luminosity of 22~$\mu\mathrm{b}^{-1}$.\
To enhance the statistics of ultra-central collisions, additional data samples were recorded 
by two dedicated triggers -- 
UCC-1 and UCC-2 -- that selected events in which the total \ET\ in the FCal at L1
  was more than 4.21~\TeV\ and 4.54~\TeV, respectively.\
The UCC-1 trigger sampled a luminosity of 45~$\mu\mathrm{b}^{-1}$ while the UCC-2 
  trigger sampled the entire luminosity of 0.49~$\mathrm{nb}^{-1}$.\
The luminosities sampled by the different triggers are listed in Table~\ref{tab:triggers}.

\begin{table}[h]
\caption{
The luminosities sampled by the triggers used in the analysis.
}
\centering
\begin{tabular}{|cc|}
\hline
Trigger      & Sampled luminosity       \\
\hline
Minimum-bias & 22~$\mu\mathrm{b}^{-1}$  \\
UCC-1        & 45~$\mu\mathrm{b}^{-1}$    \\
UCC-2        & 0.49~$\mathrm{nb}^{-1}$  \\
\hline
\end{tabular}
\label{tab:triggers}
\end{table}

In the offline analysis the $z$ coordinate of the primary vertex~\cite{Aad:2015wga} is 
  required to be within 10~cm of the nominal interaction point.\
The fraction of events containing more than one inelastic interaction
  (pile-up) is estimated to be at the level of 0.2\%.\
  The fraction varies with \fcalsum, and for ultra-central collisions it amounts to a few percent.\ 
Pile-up events were removed by exploiting the correlations between 
  the transverse energy measured in the FCal and in the ZDC as well 
  as the number of  tracks associated with the primary vertex, 
  $N^{\mathrm{rec}}_{\mathrm{ch}}$.\
  As the pile-up is very small, in a typical pile-up event the track 
  multiplicity associated with the primary vertex belongs to
  a single Pb+Pb collision, while the energy deposited in calorimeters 
  contains contributions from multiple, mostly two, collisions.\
Therefore, events with small values of $N^{\mathrm{rec}}_{\mathrm{ch}}$ 
  and large \fcalsum that differ markedly from those of the majority of \PbPb 
  collisions are removed from the analysis~\cite{HION-2012-04}.\ 
In addition, the anti-correlation between the \fcalsum and the number 
  of neutrons detected in ZDC is also used to suppress pile-up events.\ 
Events with the number of neutrons (as recorded in the ZDC) much higher 
  than the number expected from the bulk of events for a given value 
  \fcalsum are rejected.\

\begin{figure}[h!]
\centering
\includegraphics[width=0.8\textwidth]{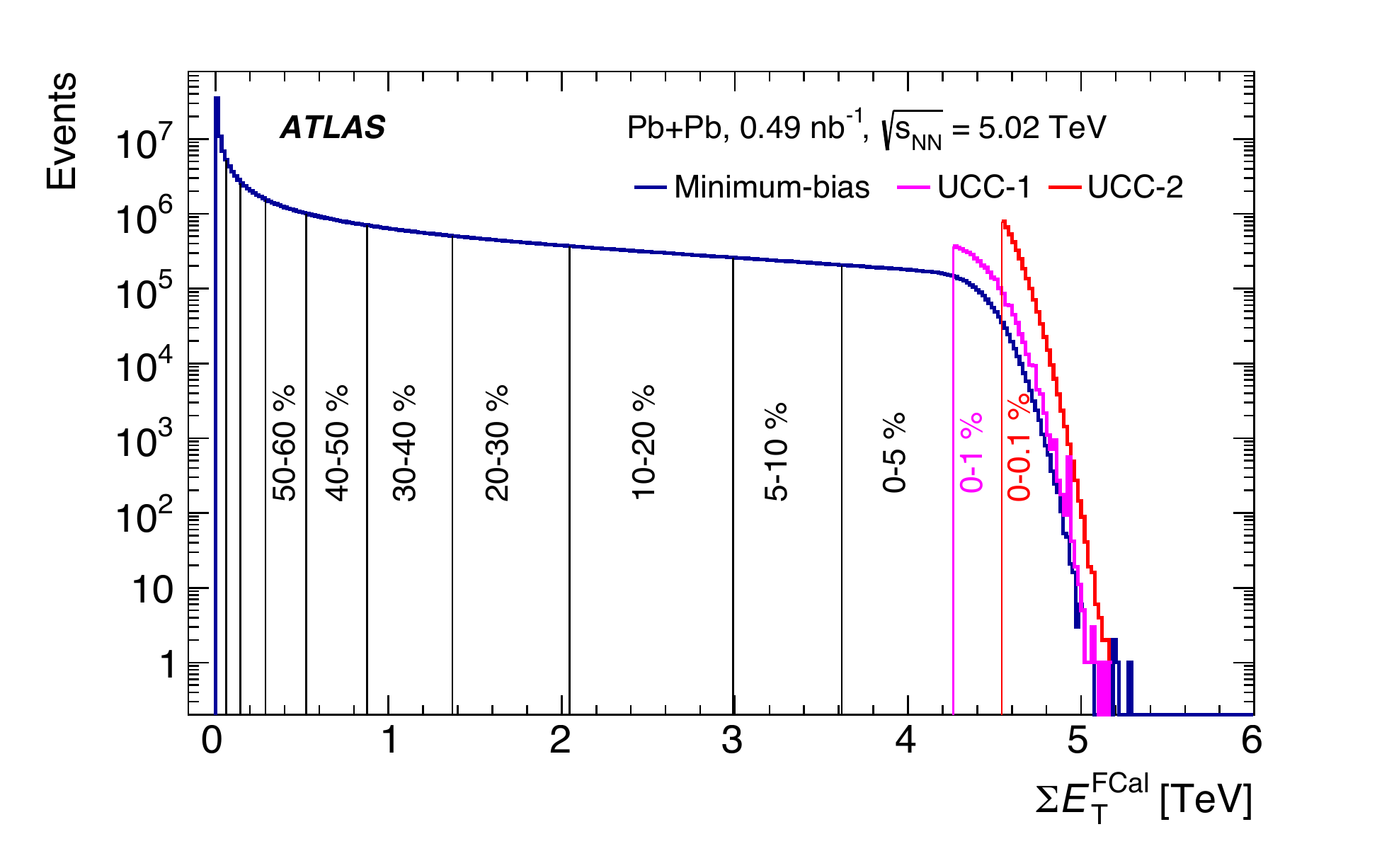}
\caption{
The $\Sigma\ET^{\mathrm{FCal}}$ distribution in \sqrtsnn~=~5.02~\TeV\ Pb+Pb data 
  for events selected by the minimum-bias trigger. 
The $\Sigma\ET^\mathrm{FCal}$ thresholds for several centrality intervals 
  are marked with vertical lines and labelled on the plot.
Also shown are the number of events over the 0--1\% and 0--0.1\%  
  centrality intervals selected by the ultra-central triggers.
}
\label{fig:centrality}
\end{figure}

The heavy-ion collision geometry is defined by its impact parameter, $b$.\ 
As the actual event-by-event impact parameter is not accessible 
  experimentally, the centrality classification is based on the 
  transverse energy measured in the forward calorimeter, \fcalsum, 
  which exhibits a strong monotonic correlation with $b$.\ 
A model based on the Monte Carlo (MC) Glauber approach~\cite{Miller:2007ri, Loizides:2014vua} 
  is used to obtain the mapping from the observed \fcalsum to the 
  primary properties, such as the number of binary nucleon--nucleon 
  interactions, $N_{\mathrm{coll}}$, or the number of nucleons 
  participating in the nuclear collision, $N_{\mathrm{part}}$, for 
  each centrality interval.\ 
The Glauber model also provides a correspondence between the \fcalsum 
  distribution and the sampling fraction of the total inelastic Pb+Pb 
  cross-section, allowing the setting of the centrality 
  percentiles~\cite{HION-2011-05}.\  
For this analysis a selection of the 80\% most central collisions
  (i.e.\ centrality \cent{0--80}) is
  used to avoid any diffractive, photonuclear, and other inelastic processes that
  contribute significantly to very peripheral collisions (centrality \cent{80--100}).\
Additionally, the events selected by UCC-1 and UCC-2 are used only over 
  the 0--1\% and 0--0.1\% centrality intervals, respectively.\
 Figure~\ref{fig:centrality} shows the distribution of \fcalsum in the data,
  and thresholds for the selection of several centrality intervals.\ 
The correspondence of centrality intervals to $\langle N_{\mathrm{part}}\rangle$ values is provided in Table~\ref{tab:cent_def}.\

\begin{table}[h!]
\centering
\caption{The correspondence between centrality intervals used in the analysis and $\langle N_{\mathrm{part}} \rangle$ values.}
\begin{tabular}{|c|c||c|c||c|c|}
\hline
Centrality& $\langle N_{\mathrm{part}} \rangle$ & Centrality& $\langle N_{\mathrm{part}} \rangle$ & Centrality& $\langle N_{\mathrm{part}} \rangle$ \\
\hline
0--0.1\% & 406.6 $\pm$ 1.3& 10--20\% & 264.1 $\pm$ 2.9 & 50--60\% & 53.9 $\pm$ 2.0\\
0--1\% & 402.9 $\pm$ 1.5& 20--30\% & 189.2 $\pm$ 2.8 & 60--70\% & 30.6 $\pm$ 1.5\\
0--5\% & 384.5 $\pm$ 1.9& 30--40\% & 131.4 $\pm$ 2.6& 70--80\% & 15.4 $\pm$ 1.0\\
5--10\% & 333.1 $\pm$ 2.7& 40--50\%&  87.0 $\pm$ 2.4 &     & \\
\hline
\end{tabular}
\label{tab:cent_def}
\end{table}

In order to study the performance of the ATLAS detector,
  a minimum-bias sample of 4M Pb+Pb MC events was generated 
  using version 1.38b of HIJING~\cite{PhysRevD.44.3501}.\
The effect of flow was added after the 
  generation using an ``afterburner''~\cite{Jia:2013tja}
  procedure in which the \pT, $\eta$ and centrality dependence 
  of the \vn, as measured in the $\sqrtsnn=2.76~\TeV$ Pb+Pb  
  data~\cite{HION-2011-01}, is implemented
  by artificially rearranging the $\phi$ positions of the generated particles.\ 
The generated sample was passed through a full simulation of the ATLAS
  detector using \GEANT\!4~\cite{Agostinelli:2002hh}, 
  and the simulated events are reconstructed using the same algorithms as used for real data.

Charged-particle tracks are reconstructed from the signals in the ID. \
A reconstruction procedure developed for tracking in dense environments in \pp collisions, and optimized for heavy-ion collisions, was used for this purpose~\cite{ATL-PHYS-PUB-2015-006}.\
In the analysis the set of reconstructed tracks is filtered using 
   several selection criteria.\
The tracks are required to have $\pT>0.5~\GeV$, $|\eta|<2.5$, at least
   two pixel hits, with the additional requirement of a hit in the first
   pixel layer when one is expected,
   at least eight SCT hits, and at most one missing hit
   in the SCT.\
 A hit is expected if the extrapolated track crosses an active
 region of a pixel module that has not been disabled, and a hit is said to be missing when it is expected but not found.\
In addition, the transverse ($d_0$) and longitudinal ($z_0\sin\theta$)
   impact parameters of the track relative to the vertex are required
   to be less than 1~mm.
The track-fit quality parameter $\chi^2/$ndof is required
   to be less than 6.

The MC sample is used to determine the track-reconstruction efficiency
  as a function of \pT, $\eta$ and centrality, $\epsilon(\pT,$  $\eta, \mathrm{centrality})$.\
 The efficiency is defined as the fraction of primary~\cite{Aad:2015wga} charged particles matched to reconstructed tracks.\
 The matching criterion is that the weighted fraction of hits in a reconstructed track originating from a given generated particle is above 30\%.\
  Different weights are assigned to pixel, SCT and TRT signals to be more robust against fake tracks, which are defined below.\ 
 At mid-rapidity ($|\eta|<1$) and for events with centrality $<5\%$, the reconstruction efficiency is $\sim$60\% 
   at low \pT\ and increases to $\sim$75\% at higher \pT.\
For $|\eta|>1$ the efficiency decreases to about 40--60\% depending on the \pt and centrality.\ 
The reconstruction efficiency depends weakly on the centrality for low-\pT\ tracks, 
   for which it is smaller in the most central events by about 5\% 
   as compared to mid-central and peripheral collisions. 
For tracks with \pT > 1~\GeV\ the dependence on centrality is less than 1\%.

The fraction of tracks that are not matched to primary,
   generated MC particles or are produced from random combinations of hits in the ID,
   both referred to as ``fake tracks'', is found to depend significantly on $\eta$.\ 
For $|\eta|<1$,  it is $\sim$10\% for low-\pT\ tracks in the most
  central 5\% \PbPb events,
  and about 5\% for more peripheral collisions.\
In the forward part of the detector, 
   especially for $1<|\eta|<2$ where detector services reside, 
   the fake rate can reach 18\% for low \pT tracks in the most central collisions.\
The fake rate drops rapidly for higher \pT\ and also decreases gradually
   towards more peripheral collisions. For $\pT\ > 10~\GeV$ and 0--5\% centrality it rises to about 5\%. 
 
\section{Analysis procedure}
\label{sec:analysis}
Three analysis techniques are used to determine the flow harmonics:
  the two-particle correlation method, which 
  uses only the information from the tracking detectors, 
  and the scalar-product and event-plane methods, which also 
  use information from the FCal.
  
In all approaches the differential flow harmonics are first obtained
  in narrow intervals of \pT, $\eta$ and centrality.\
Integrated quantities are obtained by 
  taking into account the track reconstruction efficiency, $\epsilon$,  
  and fake rate, $f$.\
A \pT-, $\eta$- and centrality-dependent weight factor $w = (1-f)/\epsilon$
  is applied to each track in the 2PC measurement and to scale each bin of
  the differential $\vn$ distributions in the SP and EP methods.

All analysis methods utilize the minimum-bias 
  sample of 22~$\mub^{-1}$.\
In addition, the SP and EP analyses use the ultra-central samples of 45~$\mub^{-1}$ and
  0.49~$\mathrm{nb}^{-1}$.

\subsection{Two-particle correlation analysis}\label{sec:2pc}
The \tpc method has been used extensively by ATLAS for correlation 
measurements~\cite{HION-2011-01,HION-2012-10,HION-2012-13,HION-2013-04,HION-2014-03,HION-2015-09,Aaboud:2016yar}.\
In the \tpc method, the distribution of particle pairs
   in relative azimuthal angle $\dphia=\phia-\phib$ and pseudorapidity
   separation $\deta=\etaaa-\etab$ is measured.\
Here the labels $a$ and $b$ denote the two particles used to 
   make the pair.\
They are conventionally called the ``trigger'' and ``associated''
   particles, respectively.\
In this analysis, the two particles are charged particles reconstructed by 
   the ATLAS tracking system, over the full azimuth and $|\eta|<2.5$,
   resulting in a pair-acceptance coverage of $\pm5.0$ 
   units in \deta.

In order to account for the detector acceptance effects, the correlation
  is constructed from the ratio of the distribution in which the 
  trigger and associated particles are taken from the same event
  to the distribution in which the trigger and associated particles 
  are taken from two different events.\ 
These two distributions are referred to as the “same-event” ($S$) 
  or “foreground” distribution and the “mixed-event” 
  or “background” ($B$) distribution, respectively, and the ratio is written as:
\begin{eqnarray*} 
\ctwo =\frac{S(\dphi,\deta)}{B(\dphi,\deta)}.
\end{eqnarray*}

The same-event distribution includes both the physical correlations 
   and correlations arising from detector acceptance effects.\
On the other hand, the mixed-event distribution reflects only the 
   effects of detector inefficiencies and non-uniformity, but contains 
   no physical correlations.\
To ensure that the acceptance effects in the $B$ distribution match closely those
   in the $S$ distribution, the $B$ distribution is constructed from 
   particles from two different events that have similar multiplicity and $z$-vertex.\
Furthermore, in order to account for the effects of tracking efficiency 
   $\epsilon(\pt,\eta)$, and fakes $f(\pt,\eta)$, each pair is weighted by
\begin{eqnarray*} 
w_{a,b} = \frac{ (1-f(\pta,\etaaa))(1-f(\ptb,\etab))}{\epsilon(\pta,\etaaa)\epsilon(\ptb,\etab)}
\end{eqnarray*}
   for $S$ and $B$.
In the ratio $C$, the acceptance effects largely
   cancel out and only the physical correlations remain~\cite{Adare:2008ae}.\
Typically, the two-particle correlations are used only to study the shape
   of the correlations in \dphi, and are conveniently normalized.\
In this \papertype, the normalization of \ctwo is chosen such that the
  \dphi-averaged value of \ctwo is unity for $|\deta|>2$.

\begin{figure}[h!]
\includegraphics[width=1.0\linewidth]{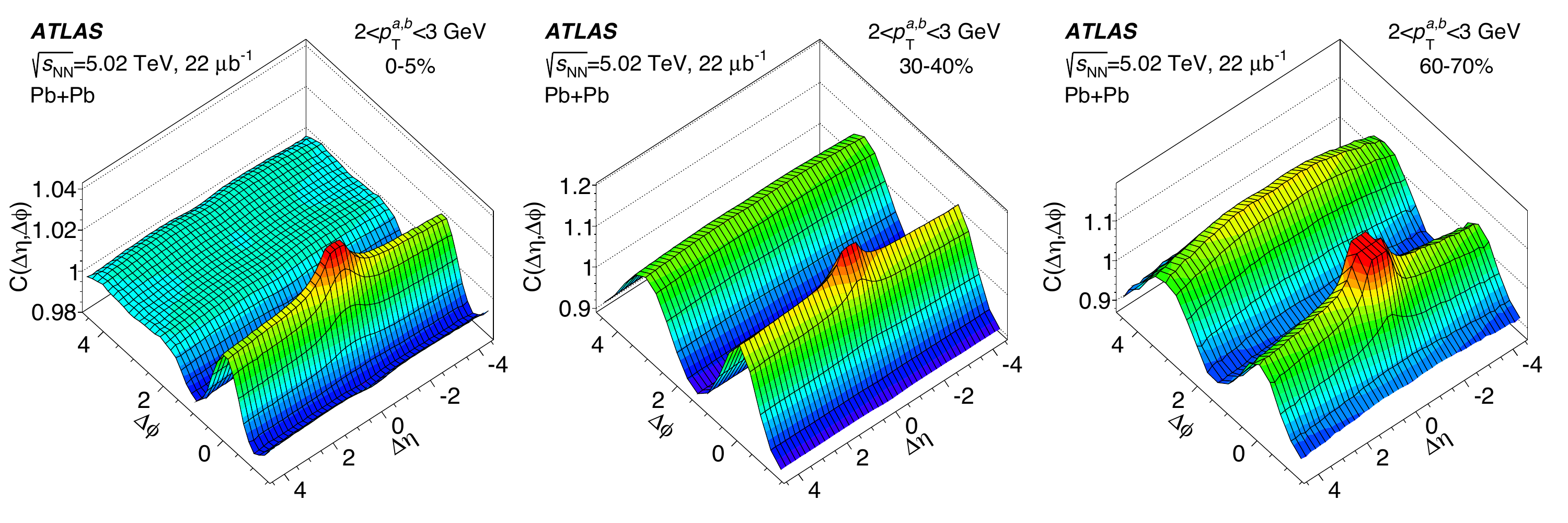}
\caption{
Two-particle correlation functions \ctwo in 5.02~\TeV\ \PbPb collisions
   for 2~<~\ptab<~3~\GeV.
The left, middle and right panels correspond to the \cent{0--5}, \cent{30--40}
   and \cent{60--70} centrality classes, respectively.\
The distributions are truncated to suppress the peak at 
   $\deta = \dphi = 0$ to show the long-range correlations in 
   greater detail.
\label{fig:2d_corrs}}
\end{figure}

Figure\,\ref{fig:2d_corrs} shows \ctwo for several centrality intervals for 
   2~<~\ptab<~3~\GeV.\
In all cases a peak is seen in the correlation at $(\deta,\dphi)\sim(0,0)$.\
This ``near-side'' peak arises from a range of sources including resonance decays, 
   Hanbury Brown and Twiss (HBT) correlations~\cite{Lisa:2005dd} and 
   jet fragmentation~\cite{HION-2015-05}.\
   The long-range (large $|\deta|$) correlations are the result of the global anisotropy
   of the event and are the focus of the study in this \papertype.

To investigate the \dphi dependence of the long-range (|\deta|~>~2)
  correlation  in more detail, the projection on to the \dphi axis is
  constructed as follows:
\begin{eqnarray*} 
\ctwophi = \frac{\int_{2}^{5} \text{d}|\deta| \; S(\dphi,|\deta|)}{\int_{2}^{5}
          \text{d}|\deta| \; B(\dphi,|\deta|)}\equiv\frac{S(\dphi)}{B(\dphi)}.
\end{eqnarray*}
The |\deta|~>~2 requirement is imposed to reject the near-side jet peak and focus 
   on the long-range features of the correlation functions.

In a similar fashion to the single-particle distribution (Eq.~(\ref{eq:single})), the
   2PC can be expanded as a Fourier series:
\begin{eqnarray}
\ctwophi=C_{0}\left(1+\Sigma_{n=1}^{\infty}\vnn(\pta,\ptb)\cos(n\dphi)\right),
\label{eq:2pc_fourier}
\end{eqnarray}
where the \vnn are the Fourier coefficients of the 2PC, and $C_0$ is its
  average value.\
If the two-particle distribution is simply the product of two
  single-particle distributions, then it can be shown that 
  the Fourier coefficients of the 2PC factorize as~\cite{Adare:2008ae}:
\begin{equation}
\vnn(\pta,\ptb)=\vn(\pta)\vn(\ptb).\
\label{eq:factorization}
\end{equation}
In Ref.~\cite{HION-2011-01} it was demonstrated that the factorization of \vnn, given by Eq.\,(\ref{eq:factorization}), is valid in central and mid-central Pb+Pb collisions at \sNN = 2.76 GeV as long as one of the correlated particles is from a low \pT range. A breakdown of the factorization is expected when the anisotropy does not arise from flow, e.g. in peripheral collisions at high \pT.
The factorization is also expected to break when the $\eta$ separation between 
   the particles is small, and short-range correlations dominate~\cite{HION-2011-01}.\
However, the |\deta|~>~2 requirement removes most such short-range correlations.\
In the phase-space region where Eq.\,(\ref{eq:factorization}) holds, the \vn(\ptb)
   can be evaluated from the measured \vnn as:
\begin{equation}
\vn(\ptb)=\frac{\vnn(\pta,\ptb)}{\vn(\pta)}=\frac{\vnn(\pta,\ptb)}{\sqrt{\vnn(\pta,\pta)}}\, ,
\label{eq:factorization_2}
\end{equation}
where $\vnn(\pta,\pta)=\vn^2(\pta)$ is used in the denominator.\
In this analysis, for most of the \tpc results the \vn(\ptb) will be evaluated 
   using Eq.\,(\ref{eq:factorization_2}) with 0.5~<~\pta<~5.0~\GeV.\
The lower cut-off of 0.5~\GeV\ on \pta is the lower limit of \pT measurements in this \papertype.\ 
The upper cut-off on \pta is chosen to exclude high-\pt\ particles, which 
 predominantly come from jets and are not expected to obey Eq.\,(\ref{eq:factorization_2}).

\begin{figure}[h!]
\includegraphics[width=1.0\linewidth]{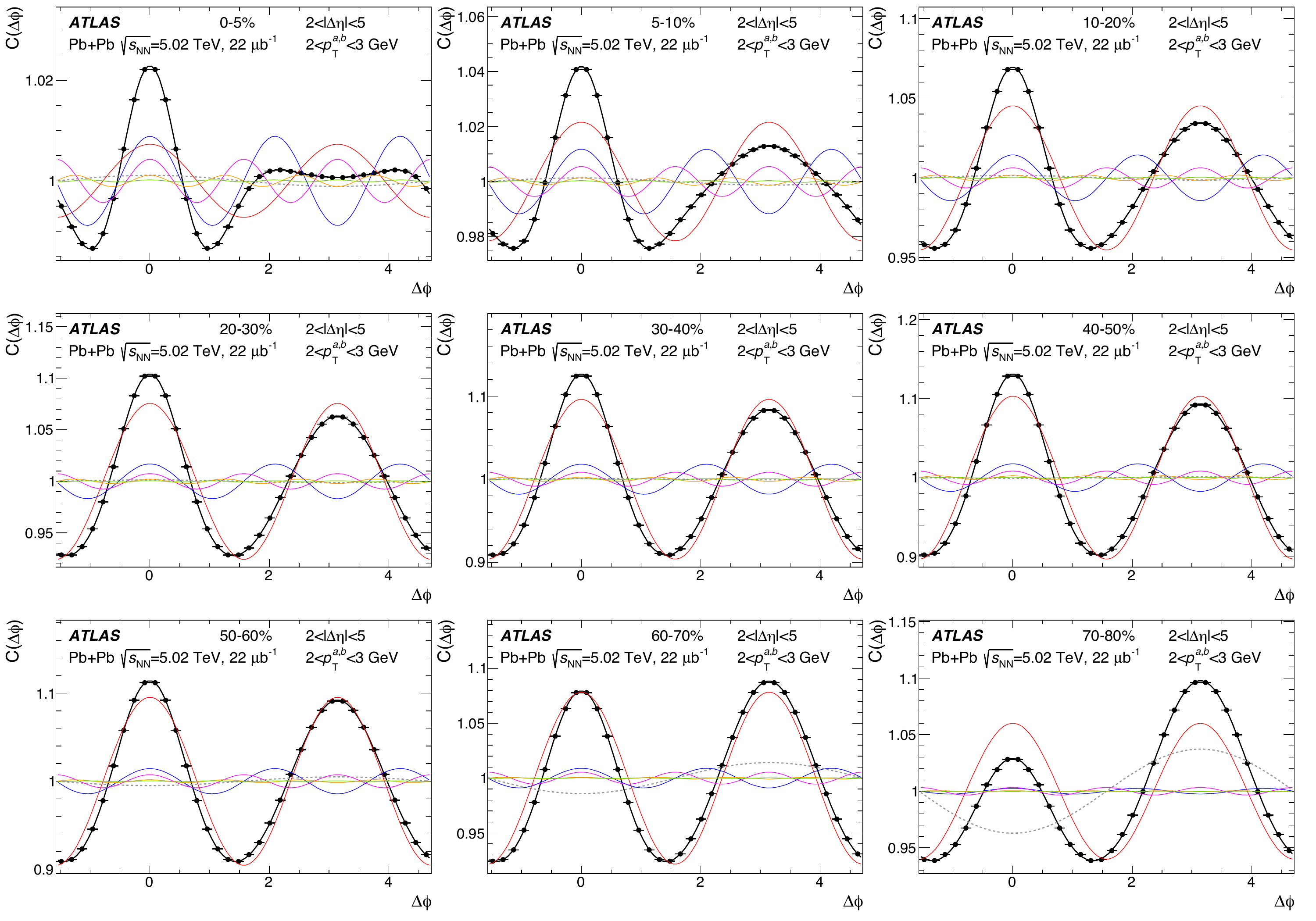}
\caption{
One-dimensional two-particle correlation functions \ctwophi in 5.02~\TeV\
   \PbPb collisions for 2~<~\ptab~<~3~\GeV\ (points).
The solid-black line indicates a fit to Eq.\,(\ref{eq:2pc_fourier})
   containing harmonics \vnn up to $n=6$.
The dashed grey line shows the contribution of the \voo.
The contributions of the \vtt--\vsisi are indicated by the coloured lines (\vtt - red,
   \vthth  - blue, \vff - magenta, \vfifi - orange, \vsisi - green), and
   can be identified by the number of peaks that they have.
Each panel corresponds to a different centrality class.
The $y$-axis range for the different panels is different. 
\label{fig:1d_corrs}}
\end{figure}

Figure\,\ref{fig:1d_corrs} shows one-dimensional 2PCs as a function of \dphi for $2<\ptab<3~\GeV$
   and for several different centrality intervals.\
The correlations are normalized to have a mean value 
   ($C_{0}$ in Eq.\,(\ref{eq:2pc_fourier})) of 1.0.\
The continuous line in Figure~\ref{fig:1d_corrs} is a Fourier fit to the correlation 
   (Eq.\,(\ref{eq:2pc_fourier}))
   that includes harmonics up to $n=6$.\
The contribution of the individual $v_{n,n}$ are also shown.\
The modulation in the correlation about its mean value is smallest in the most central
   events (top left panel) and increases towards mid-central events, reaching
   a maximum in the \cent{40--50} centrality interval and then decreases.\
In central collisions, the $v_{2,2}$--$v_{4,4}$ are of comparable magnitude.\
But for other centralities, where the average collision geometry is elongated, 
   the $v_{2,2}$ is significantly larger than the other $v_{n,n}$ for $n\geq3$.\ 
In the central events the ``away-side'' peak at $\dphi\sim\pi$ is also much broader because 
  all the significant harmonics are of similar magnitude, while in mid-central events 
   the near-side and away-side peaks are quite symmetric as the $v_{2,2}$ dominates.
In central and mid-central events, the near-side peak is larger than the away-side peak. 
However, for the \cent{60--70} and more peripheral centralities, 
   the away-side peak becomes larger due to the 
   presence of a large negative $v_{1,1}$ component. 
This negative $v_{1,1}$ component in the peripheral 2PCs arises
   largely from dijets: 
   while the near-side jet peak is rejected by the $|\deta|$~>~2 requirement, 
   the ``away-side jet'' correlation that arises from back-to-back jets 
   and contributes at $\dphi=\pi$,
   cannot be rejected entirely as its position varies in $|\deta|$ from event to event.
In the peripheral multiplicity intervals, the away-side jet significantly affects the \tpc.\
It produces a large negative $v_{1,1}$ and also affects 
   the other harmonics by adding alternately positive and negative contributions to \vnn harmonics of even and odd order, respectively.\
In peripheral events the \vnn are strongly biased by dijets especially 
   at higher \pt.\ 
The presence of the jets also results in the breakdown of the factorization 
   relation (Eq.\,(\ref{eq:factorization})).

\subsection{Scalar Product and Event Plane analysis}
The SP method was introduced
  by the STAR Collaboration~\cite{Adler:2002pu} 
  and is further discussed in Ref.~\cite{Voloshin:2008dg}.\  
The SP method is very similar to the Event Plane method (EP)
  widely used in earlier analyses~\cite{HION-2011-05,HION-2011-01}.\
It is superior to the EP
  as $v_{n}\{\mathrm{SP}\}$ is an estimator of $\sqrt{\langle \vn^{2}\rangle}$, 
  independent of the detector resolution and acceptance, whereas $v_{n}\{\mathrm{EP}\}$ produces 
  a detector-dependent estimate of \vn that lies between
  $\langle \vn \rangle$ and $\sqrt{\langle \vn^{2}\rangle}$~\cite{Luzum:2012da}.

Both the SP and EP method use flow vectors $Q_{n}$ and $q_{n,j}$ defined as: 
  \begin{equation}
    Q_{n} = |Q_{n}|\text{e}^{\text{i}n\Psi_{n}} = 
    \frac{1}{M}\sum_{j=1,M}q_{n,j} = 
    \frac{1}{M}\sum_{j=1,M}w_j \text{e}^{\text{i}n\phi_{j}},
    \label{eq:FlowVector}
  \end{equation}
  where the sum runs over $M$ particles in a single event.\
The  $\phi_{j}$ is the particle azimuthal angle and $n$ is the harmonic order.\ 
In this analysis the flow vectors are established separately
  for the two sides of the FCal and are denoted $Q_n^{N|P}$, 
  where the N and P correspond to $\eta < 0$  and  $\eta >0$ sides, respectively.\
The sum in Eq.~(\ref{eq:FlowVector}) in this case runs
  over the calorimeter towers of  approximate granularity 
  $\Delta\eta \times \Delta\phi = 0.1 \times 0.1$ and the weights $w_i$ 
  are the transverse energies \ET\ measured in the FCal towers.\
The flow vectors are also calculated using charged-particle tracks. In this case the sum in Eq.~(\ref{eq:FlowVector})  is over tracks and $w_{j}$ is obtained as the MC tracking weight ($(1-f)/\epsilon$) multiplied by a factor that depends on azimuthal angle to correct for non-uniformity in the azimuthal-angle distribution of reconstructed tracks.\
This latter factor is obtained run-by-run from the data as the average track multiplicity in one $\eta$ slice of 0.1 divided by the multiplicity in the narrow $\Delta\eta \times \Delta\phi = 0.1 \times 0.1$ interval.

\begin{figure}[!htb]
  \centering
  \includegraphics[width=\textwidth]{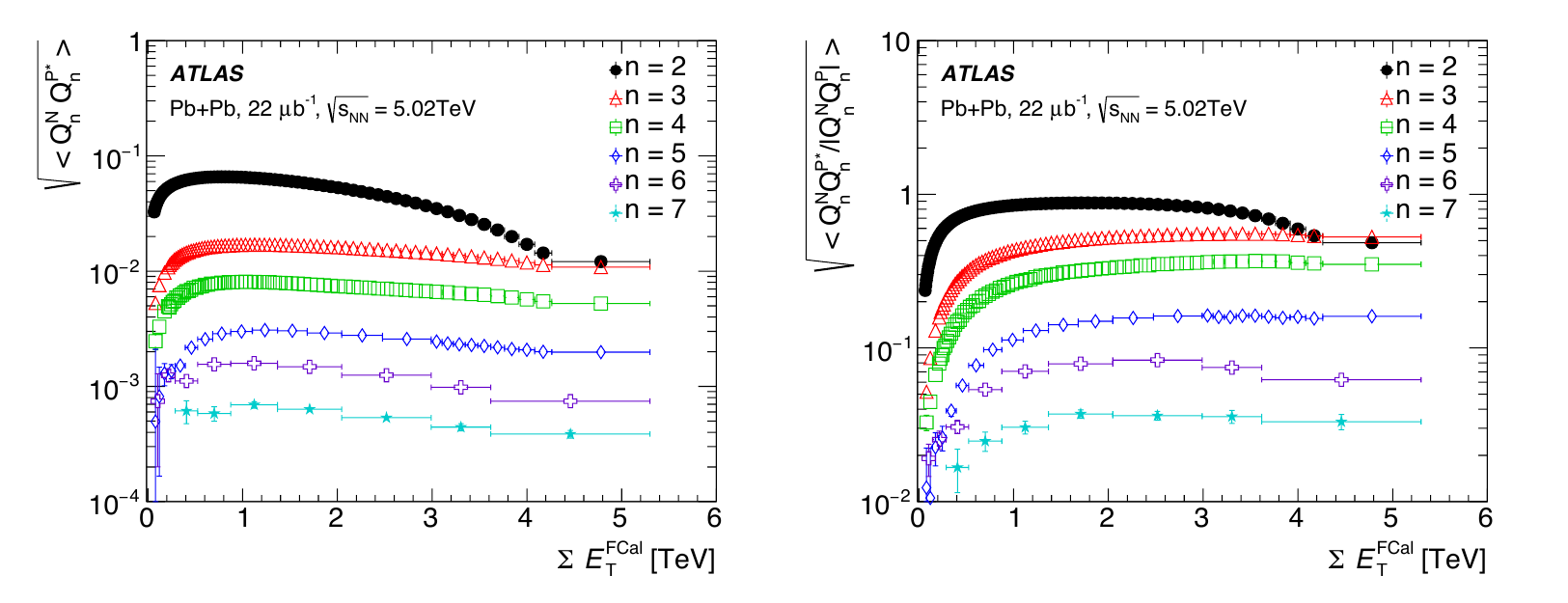}
  \caption{The dependence of the correction factor in the SP method,
    $\sqrt{\langle Q_{n}^{N}Q_{n}^{P*}\rangle}$ (left panel), and EP method, $\sqrt{\Big\langle\frac{Q_{n}^{N}Q_{n}^{P*}}{|Q_{n}^{N}||Q_{n}^{P}|}\Big\rangle}$ (right panel), for all measured
    harmonics as a function of \fcalsum binned according to the
    centrality bins definition.}
  \label{fig:resolSP}
\end{figure}

The main idea of the SP method is to correlate single-track unit flow vectors with the flow vector of all particles measured in the FCal region ($3.2 < |\eta| < 4.9$).\
Therefore, the SP method differs from the two-particle correlation method, 
in which each single track is correlated with all tracks of $|\Delta\eta|>2$ in the event.\
The values of $\vn$ in this analysis are obtained as:
  \begin{equation}
    v_{n}\{\mathrm{SP}\} = 
    Re \frac{\langle q_{n,j}Q_{n}^{N|P*} \rangle}{\sqrt{\langle
        Q_{n}^{N}Q_{n}^{P*}\rangle}} = 
    \frac{\langle |q_{n,j}||Q_{n}^{N|P}|\cos[n(\phi_{j} -
      \Psi_{n}^{N|P})] \rangle}
    {\sqrt{\langle |Q_{n}^{N}||Q_{n}^{P}|\cos[n(\Psi_{n}^{N} - \Psi_{n}^{P})] \rangle}},
    \label{eq:vnSP}
  \end{equation}
  where $q_{n,j}$ is the flow vector obtained for a small ($\eta,
  \pt$) interval  (typically 0.1 in $\eta$ and 0.1~\GeV\ in \pT below 5~\GeV\
  and 1~\GeV\ at higher \pT) using tracks,  
  $Q_{n}^{N|P}$ is the flow vector obtained using
  either the N or P side of the FCal, chosen so
  that the $\eta$ gap between the $q_{n,j}$ and
  $Q_{n}$ is maximized,
  the * denotes complex conjugation, the $\Psi_{n}$ are estimates of 
  the $n^{\text{th}}$-order reaction-plane angles (Eq.~(\ref{eq:single})) and
  the angular brackets indicate an average over all events.\ 
In the last line of Eq.~(\ref{eq:vnSP}) it is assumed that the sine terms disappear, as required from symmetry.\
The correction factor, $1/\sqrt{\langle Q_{n}^{N}Q_{n}^{P*}\rangle}$,
 (Eq.~(\ref{eq:vnSP})) depends on the harmonic order
 and \fcalsum\ as shown in the left panel of Figure~\ref{fig:resolSP}.\
 The event-plane angles, $\Psi_{n}$, and the $Q_{n}$ vectors, both measured in the FCal, may be biased due to non-uniform detector response.\
 As $\Psi_{n}$ varies randomly from event to event, its distribution should be uniform, and the components of the $Q_n$ vector,  $Q_{n,x} = |Q_{n,}|\mathrm{cos}(\Psi_{n})$  and $Q_{n,y} = |Q_{n}|\mathrm{sin}(\Psi_{n})$, should be zero when averaged over many events.\
This is achieved by the following procedure. In its first step, non-zero offsets of the mean of raw flow vector coordinates are removed for each run:
$Q_{n,i} =Q^{\mathrm{raw}}_{n,i}  - \langle Q^{\mathrm{raw}}_{n,i}\rangle$  where $i=x,y$ and $\langle Q^{\mathrm{raw}}_{n}\rangle$ is the mean 
 calculated for each run.\ 
 However, even after this correction, residual higher-order non-uniformities persist, indicated by non-zero values of $\langle Q_{n,x}Q_{n,y}\rangle$.\
 These are removed  by rotating the $Q_{n}$ vector so that the corrected $Q_{n}$ vector has no skew  ($\langle Q_{n,x}^{2}\rangle= \langle Q_{n,y}^{2}\rangle$;  $\langle Q_{n,x} Q_{n,y}\rangle = 0$) and the distributions of the resulting EP angles, $\Psi_{n}$, are uniform~\cite{ATLAS-CONF-2017-008}.

In the Event Plane analysis the reference $Q_{n}$ vectors
  are normalized to unity, $Q_n^{N|P} \to Q_n^{N|P}/|Q_n^{N|P}|$,
  before using them in Eq.~(\ref{eq:vnSP}). 
So the  \vn estimate is obtained as:
\begin{equation}
  \vn\{\mathrm{EP}\}  = Re \frac{\Big\langle
    q_{n,j}\frac{Q_{n}^{N|P*}}{|Q_{n}^{N|P}|}\Big\rangle}{\sqrt{\Big\langle\frac{Q_{n}^{N}}{|Q_{n}^{N}|}\frac{Q_{n}^{P*}}{|Q_{n}^{P}|}\Big\rangle}}
  = \frac{\langle\cos[n(\phi_{j} - \Psi_{n}^{N|P})]
    \rangle}{\sqrt{\langle\cos[n(\Psi_{n}^{N} - \Psi_{n}^{P})] \rangle}}.
  \label{eq:vnEP}
\end{equation}

The denominator of Eq.~(\ref{eq:vnEP}), shown in the right panel of Figure~\ref{fig:resolSP}, can be thought of as a resolution.\
It is distinct for each harmonic and depends on \fcalsum.

In this analysis the EP method is used only for the purpose 
  of a direct comparison with the results obtained in Run 1~\cite{HION-2011-01},
  in which only the EP method was used.

The analysis is performed in intervals of centrality.  
  The \vn\  values are obtained in narrow bins of \pT\ and $\eta$,
  which are summed, taking into account tracking efficiency and fake rate, 
  to obtain the integrated results.

\FloatBarrier

A detailed study based on a HIJING~\cite{PhysRevD.44.3501} Monte Carlo 
  sample showed a difference for the most central events between the 
  \vn obtained with generated particles and the \vn obtained with 
  reconstructed tracks (the ``MC closure test'').\ 
The discrepancies are due to the presence of fake tracks, which at 
  low \pT\ distort the \vn measurements, and a tracking inefficiency 
  in the event-plane direction due to increased detector occupancy 
  resulting from the flow phenomenon itself, which lowers the 
  measured \vn values.\
The study based on the $d_0$ distribution also showed that the fake-track 
  rates are overestimated in  MC simulation as compared to the data.\
This disagreement was removed by weighting MC tracks so that the 
  $d_0$-distribution tails ($2 < |d_0| < 10$ mm)  match those in data, 
  following the procedure described in Ref.~\cite{STDM-2010-01}.\  
It was observed that the contribution of fakes to the ``MC non-closure'' 
  is significant only for events with centrality $<$~30\% and at low \pT, 
  which is the region where the fake rate is the largest.\ 
In this modified MC sample, the relative differences between values 
  of the \vn measured with generated particles and reconstructed tracks 
  are used as corrections to account for both effects; the fakes and the 
  $\Psi_{n}$-dependent inefficiency.\
Corrections are at most 5--10\%.\ 
For example, for \vtwo in the 0--5\% centrality interval,
the correction is as large as 7\% at low \pT and becomes negligible above \pT $> 2$~\GeV.\
Corrections of a similar magnitude are obtained for higher-order harmonics.

\section{Systematic uncertainties}
\label{sec:systematics}
The systematic uncertainties of the measured \vn are evaluated by 
   varying several aspects of the analysis.\
As the EP and SP results are subject to the same uncertainty sources, the uncertainty 
  values are of the similar magnitude and are not discussed separately.\
Similarly, some of the uncertainty sources are common to the SP/EP and the \tpc methods and are discussed together.\
The uncertainties for two representative \pT\ intervals are summarized 
  in the Tables~\ref{tab:unc2PC} and~\ref{tab:uncSP} 
  for the \tpc and SP/EP methods, respectively.\ 
In the discussion below, other \pT\ ranges are referred to
  if uncertainties are significantly higher than in the \pT\ ranges shown in the tables.\
The following sources of uncertainty are considered:

\begin{itemize}
\item \textbf{Track selection:} 
The tracking selection requirements control the relative contribution of 
   genuine charged particles and fake tracks entering the analysis.\
The stability of the results to the track selection is evaluated by 
   varying the requirements imposed on the reconstructed tracks.\
Two sets of variations are used.\
In the first case the required number of pixel and SCT hits on the 
  reconstructed track are relaxed to one and six, respectively.\
Additionally, the requirements on the transverse and longitudinal 
  impact parameters of the track are relaxed to 1.5~mm.\
In the second case, the track selection is based on requirements 
used for the baseline measurement, but the transverse and longitudinal 
impact parameters of the track are restricted to 0.5~mm.\
For each variation, the entire analysis is repeated including the 
   evaluation of the corresponding efficiencies and fake rates.\
The fake rate is largest at the lowest \pT (0.5 GeV) and for the most central events, and consequently the variation in the \vn values obtained from this procedure is largest, typically 10\%, in this region of phase space.

\item \textbf{Tracking efficiency:}
As mentioned above, the tracks are weighted by a factor $(1-f)/\epsilon(\pt,\eta)$
   when calculating the \vn to account for the effects of the 
   tracking efficiency.\
Uncertainties in the efficiency, resulting e.g.\ from an uncertainty in the amount of 
   detector material, need to be propagated into the measured \vn~\cite{Aad:2016mok}.\
This uncertainty is evaluated by varying the efficiency up and down 
   within its uncertainties in a \pT-dependent manner and re-evaluating the \vn.\
This contribution to the overall uncertainty is very small and 
   amounts to less than 1\% on average.\
This is because the change of efficiency largely cancels out in the differential
  $\vn(\pT)$ measurement, and for \vn integrated over \pT
  the low-\pT particles dominate the measurement.\
It does not change significantly with centrality nor with the 
   order of harmonics.

\item \textbf{Centrality determination:} 
An uncertainty in the flow harmonics comes
   from the uncertainty in the fraction of the total inelastic
   cross-section accepted by the trigger and the event selection
   criteria, which was estimated to be at a level of 1\%.\
The \vn uncertainty is evaluated by repeating the analysis with the
modified centrality selections on the \fcalsum\ distribution shown in Fig.~\ref{fig:centrality}, that give the $\pm$ 1\% uncertainty 
   in the sampled fraction of the cross-section~\cite{HION-2011-05}.\
The changes in the \vn are largest in the peripheral-centrality 
   intervals, for which the bin definitions are 
   significantly changed when remapping the centralities.\
For \vtwo, a change of $\sim$0.8\% (2PC) and
   $\sim$1\% (SP)  is also observed in the most central events.\
This is because the \vtwo changes rapidly with centrality
   in central events, so slight variations in the centrality 
   definition result in significant change in \vtwo.\
For \vthree this uncertainty varies from less than 0.5\% over the
   \cent{0--50} centrality range to $\sim$5\% in the \cent{70--80} centrality interval.\
For the higher-order harmonics, $n>3$, the  uncertainty is less 
   than 0.5\% over the \cent{0--50} centrality range and increases
   to about 2\% for more peripheral bins.\
The variation in the \vn when using these alternative centrality
   definitions is taken as a systematic uncertainty.\
To limit  the statistical instability of \vsix and \vseven in uncertainty estimation, the variations for this measurement were determined over a wide range of \pT = 0.5--60 \GeV.

\item \textbf{MC corrections:} 
To assess the uncertainty related to the MC corrections the closure test is repeated with the
  two selections of tracks described in the ``track selection'' paragraph.\
Differences between the correction factors obtained with loose,
  nominal and tight tracking selections are compared.\
The difference between them is largest at low \pT and central
  events and amounts typically to a few percent.\
It is negligibly small above 2 \GeV.\
The larger of the two differences (between the nominal
  and loose tracking selections) is used as an uncertainty estimate.

\item \textbf{Residual sine term:} 
The ability of the detector to measure small \vn signals can be 
   quantified by comparing the value of the \vn calculated as 
   the real part of the flow vector product (SP) in Eq.~\eqref{eq:vnSP}  
with its imaginary part.\
The ratio $Im(SP)/\vn$ is taken as a contribution to the 
   systematic uncertainty.\
The contribution from this source is  $\sim$1\% in most of 
   the phase space, while for the higher harmonics ($n=6, 7$) 
   it can reach 20\% in the most central collisions.\
This uncertainty is only relevant for the \vn values measured by the 
   EP and SP methods.

\item \textbf{Variation of FCal acceptance in the $Q_n^{N|P}$ estimation:}
In order to quantify an uncertainty arising from the FCal acceptance in the
   $Q_n^{N|P}$ estimation, 
   \vn harmonics are compared for two distinct 
   FCal regions $3.2<|\eta|<4$ and $4<|\eta|<4.8$ used for  the
   determination of the reference flow vector, $Q_n$.\
The differences between the \vn values are treated as the systematic
  uncertainty, which, similarly to the $\eta$ symmetry (next paragraph), quantifies
  the ability of the detector to measure small signals.\
Accordingly, this contribution is small ($\sim$ 1\%) for \vtwo and \vthree and
  starts growing for higher-order harmonics, reaching about 27\% for \vseven.\
This uncertainty is only relevant to the \vn values measured by the 
   EP and SP methods.

\item \textbf{Detector non-uniformity:}
Due to the symmetry of the \PbPb collision system the $\vn(\eta)$ are expected to be on average symmetric in rapidity. 
Any difference between the event-averaged \vn at $\pm\eta$ arises 
   from residual detector non-uniformity.\
The difference between the \vn values measured in opposite 
   hemispheres is treated as the systematic uncertainty quantifying 
   non-perfect detector performance.\
This uncertainty is in general very low (less than 1\%) except for high-order
  harmonics:  \vfive and \vsix at high \pT and \vseven at all \pT.
This uncertainty only contributes to the \vn values measured by the 
   EP and SP methods.\
For the \tpc method, the residual non-uniformity is estimated by 
  varying the event-mixing procedure.

\item \textbf{Event-mixing:}
As explained in Section\,\ref{sec:2pc}, the \tpc analysis uses the event-mixing
   technique to estimate and correct for the detector-acceptance effects.\
Potential systematic uncertainties in the \vn due to the residual pair-acceptance 
   effects, which were not removed by the mixed events, are evaluated
   by varying the multiplicity and $z$-vertex matching criteria used to
   make the mixed-event distributions, following Ref.\,\cite{HION-2011-01}.\
The resulting uncertainty for \vtwo--\vfive is between 1--3\%, and for 
   \vsix is between 4--8\% for most of the centrality and \pt\ ranges 
   measured in this \papertype.\
However, the uncertainties for \vfour--\vsix are significantly larger for 
   $\pt\ {<}$ 0.7~\GeV, where the \vn signals are quite small and very 
   susceptible to acceptance effects, and for \vsix are correlated with 
   statistical uncertainties.\
The uncertainties are also significantly larger for \mbox{$\pt\ >$ 10~\GeV,} 
   where they are difficult to determine due to large statistical uncertainties 
   in the measurements.

\end{itemize}

\begin{table}
\caption{
The systematic uncertainties associated with the \tpc \vn measurements for
   selected intervals of \pt and for 5--10\% and 40--50\% centrality bins.\
The contributions are expressed in \%.\
The total systematic uncertainty is obtained by adding the contribution 
  of the individual sources in quadrature.
}
\begin{center}
\begin{tabular}{|c c|c c|c c|}
\hline
\multirow{2}{1.8cm}{\centering Systematic sources } & 
\multirow{2}{1.3cm}{\centering $n^{\text{th}}$ harmonic} &
                    \multicolumn{2}{|c|}{ 5--10\% } & \multicolumn{2}{|c|}{40--50\%}   \\ \cline{3-6}
                    & & 0.8--1.0~\GeV & 6--8~\GeV         &0.8--1.0~\GeV & 6--8~\GeV         \\
\hline 
\multirow{6}{1.8cm}{ \centering Track selection}     
    & \vtwo                 &  0.5   &  0.5               &  0.5    & $<\!0.5$    \\
    & \vthree               &  1     &  1                 &  0.5    &  0.5    \\
    & \vfour                &  0.5   &  $<\!0.5$              & $<\!0.5$    &  1      \\
    & \vfive                &  2     &  $<\!0.5$             &  0.5    &  5      \\
    & \vsix                 &  2     &  2                 & 2       &  2      \\
\hline
\multirow{6}{1.8cm}{ \centering Tracking efficiency}     
    & \vtwo                 &  0.1  & 0.1          & 0.1 & 0.1   \\
    & \vthree               &  0.1  & 0.1          & 0.1 & 0.1   \\
    & \vfour                &  0.1  & 0.1          & 0.1 & 0.1   \\
    & \vfive                &  0.1  & 0.1          & 0.1 & 0.3   \\
    & \vsix                 &  1    & 0.1          & 1   & 0.1   \\
\hline
\multirow{6}{1.8cm}{ \centering Centrality determination}     
    & \vtwo                 &  1    & 1                 &0.5  & 0.5        \\
    & \vthree               &  0.5  & 0.5               &0.5  & 3          \\
    & \vfour                &  0.5  & 0.5               &0.5  & 3          \\
    & \vfive                &  0.5  & 0.5               &0.5  & 3          \\
    & \vsix                 &  0.5  & 0.5               &0.5  & 3          \\
\hline
\multirow{7}{1.8cm}{ \centering MC corrections}     
    & \vtwo                 &  2   & $<\!0.5$                  & $<\!0.5$   & $<\!0.5$    \\
    & \vthree               &  2   & $<\!0.5$                  & $<\!0.5$   & $<\!0.5$    \\
    & \vfour                &  1   & $<\!0.5$                  & $<\!0.5$   & $<\!0.5$   \\
    & \vfive                &  1   & $<\!0.5$                  & 1      & 1       \\
    & \vsix                 &  3   & $<\!0.5$                  & 2      & $<\!0.5$    \\
\hline
\multirow{6}{1.8cm}{ \centering Event-mixing}     
    & \vtwo                 &  1& 1        & 1&  1      \\
    & \vthree               &  1& 3        & 1&  3      \\
    & \vfour                &  2& 6        & 1&  6      \\
    & \vfive                &  3&10        & 3& 10      \\
    & \vsix                 &  5&15        & 5& 15      \\
\hline
\end{tabular}
\end{center}
\label{tab:unc2PC}
\end{table}

\begin{table}
\caption{
The systematic uncertainties associated with the SP and EP 
   (in parentheses) \vn measurements for \vn\ in 5--10\% and 40--50\%
   centrality bins.\ 
The contributions are expressed in \%.\ 
The total systematic uncertainty is obtained by adding the contribution 
  of the individual sources in quadrature.
}
\begin{center}
\begin{tabular}{|c c|c c|c c|c c|}
\hline
\multirow{2}{1.8cm}{\centering Systematic sources } & 
\multirow{2}{1.3cm}{\centering $n^{\text{th}}$ harmonic} &
 \multicolumn{2}{|c|}{ 5--10\% } & \multicolumn{2}{|c|}{40--50\%}   \\ \cline{3-6}
								 	& & 0.8--1~\GeV & 9--10~\GeV 		&  0.8--1~\GeV & 9--10~\GeV   \\
\hline 
\multirow{6}{1.8cm}{ \centering Track selection} 	
													& \vtwo 	  &  0.5 (1) & 0.5 ($<\!0.5$) 		& $<\!0.5$ ($<\!0.5$) & $<\!0.5$ ($<\!0.5$)	  			\\
  													& \vthree 	  &  1 (1) & 1 ($<\!0.5$) 			&  0.5 (0.5) & 0.5 (0.5) 				\\
  													& \vfour	  &  0.5 (0.5) & $<\!0.5$ (0.5) 		& $<\!0.5$ ($<\!0.5$) & 1 (1)				\\
  													& \vfive 	  &  2 (1) & 0.5 ($<\!0.5$) 			& 0.6 (0.5) & 5 (4)			\\
  													& \vsix	&  \multicolumn{2}{c|}{2 (2)} 	& \multicolumn{2}{c|}{2 (2)}		 \\
  													& \vseven	   &  \multicolumn{2}{c|}{6 (6)} 	& \multicolumn{2}{c|}{4 (5)}		\\
\hline
\multirow{6}{1.8cm}{ \centering Tracking efficiency} 	
													& \vtwo 		&  \multicolumn{2}{c|}{0.1 (0.1)}	& \multicolumn{2}{c|}{0.1 (0.1)}	\\
  													& \vthree 	        &  \multicolumn{2}{c|}{0.1 (0.1)} 	& \multicolumn{2}{c|}{0.1 (0.1)}  	\\
  													& \vfour		&  \multicolumn{2}{c|}{0.1 (0.1)} 	& \multicolumn{2}{c|}{0.1 (0.1)} 			\\
  													& \vfive 		&  \multicolumn{2}{c|}{0.1 (0.1)} 	& \multicolumn{2}{c|}{0.1 (0.1)} 			\\
  													& \vsix		&  1 (1) & 0.1 (0.1) 			& 1 (1) & 0.1 (0.1)					\\
  													& \vseven		&  \multicolumn{2}{c|}{1.5 (1.5)} 		&\multicolumn{2}{c|}{1.5 (1.5)}			 \\
\hline
 \multirow{6}{1.8cm}{ \centering Centrality determination} 	
													& \vtwo 		&  0.5 (0.5) & 0.5 (0.5) 		& $<\!0.5$ ($<\!0.5$) & $<\!0.5$ ($<\!0.5$)		\\
  													& \vthree 		&  $<\!0.5$ ($<\!0.5$) & $<\!0.5$ ($<\!0.5$) 		& $<\!0.5$ ($<\!0.5$) & 0.5 (1)	 \\
  													& \vfour		&  $<\!0.5$ ($<\!0.5$) & $<\!0.5$ ($<\!0.5$) 	& 0.5 (0.5) & $<\!0.5$ ($<\!0.5$)		 \\
  													& \vfive 		&  $<\!0.5$ ($<\!0.5$) & $<\!0.5$ (0.5) 		& 1 (1) & 1 (1)		      		\\
  													& \vsix		&  \multicolumn{2}{c|}{2 (2)}	& 2 (3) & 2 (3)				\\
  													& \vseven		&  \multicolumn{2}{c|}{2 (3)}	&   \multicolumn{2}{c|}{5 (5)}		\\
\hline
\multirow{6}{1.8cm}{ \centering Residual sine term} 	
													& \vtwo 		&  0.5 (0.5) & 0.5 (0.5) 		& 0.5 (0.5) & 0.5 (0.5)		\\
  													& \vthree 		&  1 (1) & 1 (1) 		& 0.5 (1) & 0.5 (0.5)		\\
  													& \vfour		&  1 (0.5) & 1 (1) 			& 1 (0.5) & 1 (1)			\\
  													& \vfive 		&  1 (0.5) & 1 (1) 				& 1 (1) & 0.5 (0.5)				\\
  													& \vsix		&  22 (26) & 2 (1) 			& 19 (11) & 1 (3)  			\\
  													& \vseven		&  \multicolumn{2}{c|} {20 (20)} & \multicolumn{2}{c|}{17 (4)} 		\\
\hline
\multirow{7}{1.8cm}{ \centering MC corrections} 	
													& \vtwo 		&  2 (2) & $<\!0.5$ ($<\!0.5$) 		& $<\!0.5$ ($<\!0.5$) & $<\!0.5$ ($<\!0.5$)		\\
  													& \vthree 		&  2 (2) & $<\!0.5$ ($<\!0.5$) 		& $<\!0.5$ ($<\!0.5$) & $<\!0.5$ ($<\!0.5$)		\\
  													& \vfour		&  1 (1) & $<\!0.5$ ($<\!0.5$) 		&  $<\!0.5$ ($<\!0.5$) & $<\!0.5$ ($<\!0.5$)		\\
  													& \vfive 		&  1 (1) & $<\!0.5$  (0.5 ) 		& 1 (1) & 1 (0.5)		\\
  													& \vsix		&  3 (3) & $<\!0.5$ (0.5) 				& 2 (2) & 0.5 (0.5)	\\
  													& \vseven		&  - & - 		& - & -		\\
\hline
\multirow{6}{1.8cm}{ \centering  FCal response} 	
													& \vtwo 		&  $<\!0.5$ (1) & 0.5 (1) 		& $<\!0.5$ (0.5) & 1 (1)		\\
  													& \vthree 		&  0.5 (0.5) & 0.5 (1) 				& $<\!0.5$ ($<\!0.5$) & 2 (3)		 	\\
                                              								& \vfour		&  1 (2) & $<\!0.5$ ($<\!0.5$) 				& 1 (1) & 2 (2)				\\
  													& \vfive 		&  1 (1) & 3 (1)				& 4 (8) & 9 (16)					\\
  													& \vsix		&  \multicolumn{2}{c|}{3 (5)} 		& \multicolumn{2}{c|}{16 (14)} 	\\
  													& \vseven		&  \multicolumn{2}{c|}{27 (34)}	& \multicolumn{2}{c|}{20 (9)}		\\
\hline

\multirow{6}{1.8cm}{ \centering Detector non-uniformity} 	
													& \vtwo 		&  $<\!0.5$ ($<\!0.5$) & $<\!0.5$ ($<\!0.5$) 		& $<\!0.5$ ($<\!0.5$) & $<\!0.5$ ($<\!0.5$)		\\
  													& \vthree 		&  0.5 ($<\!0.5$) & $<\!0.5$ ($<\!0.5$) 		& $<\!0.5$ ($<\!0.5$) & $<\!0.5$ ($<\!0.5$)			\\
  													& \vfour		&  $<\!0.5$ (1) & $<\!0.5$ (0.5) 		& 0.5 (0.5) & $<\!0.5$ (0.5)	 	 \\
  													& \vfive 		&  2 (2) & 1 (0.5) 				& 1 (0.5) & 1 (0.5)					\\
  													& \vsix	 	&  \multicolumn{2}{c|}{8 (10)} 		& \multicolumn{2}{c|}{0.5 (2)}		\\
  													& \vseven		&  \multicolumn{2}{c|}{2 (3)}		&   \multicolumn{2}{c|}{18 (14)}		\\
\hline
\end{tabular}
\end{center}
\label{tab:uncSP}
\end{table}

\clearpage
\section{Results}
\label{sec:results}
\begin{figure}[ht!]
\includegraphics[width=0.99\linewidth]{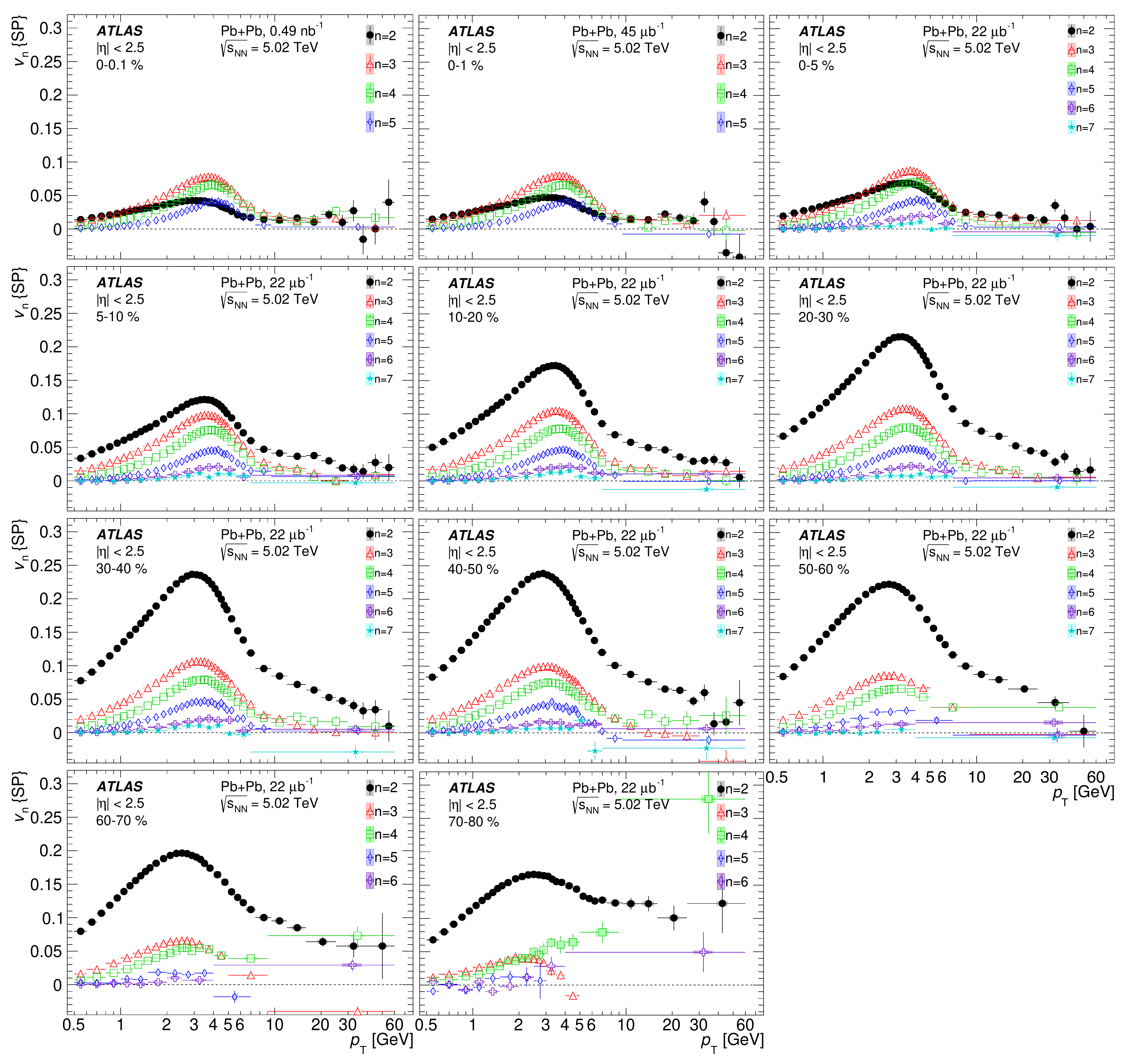}
\caption{
The \vn obtained with the SP method as a function of transverse momentum \pT, 
  integrated over $|\eta| < 2.5$ in 11 centrality intervals, from the most central 
  at the top left panel to the most peripheral at the bottom right panel.\ 
Results are averaged over the intervals indicated by horizontal error bars.\
The vertical error bars indicate statistical uncertainties; the shaded boxes indicate systematic uncertainties.
}
\label{fig:vnSP_pt}
\end{figure}

\begin{figure}[h!]
\includegraphics[width=0.99\linewidth]{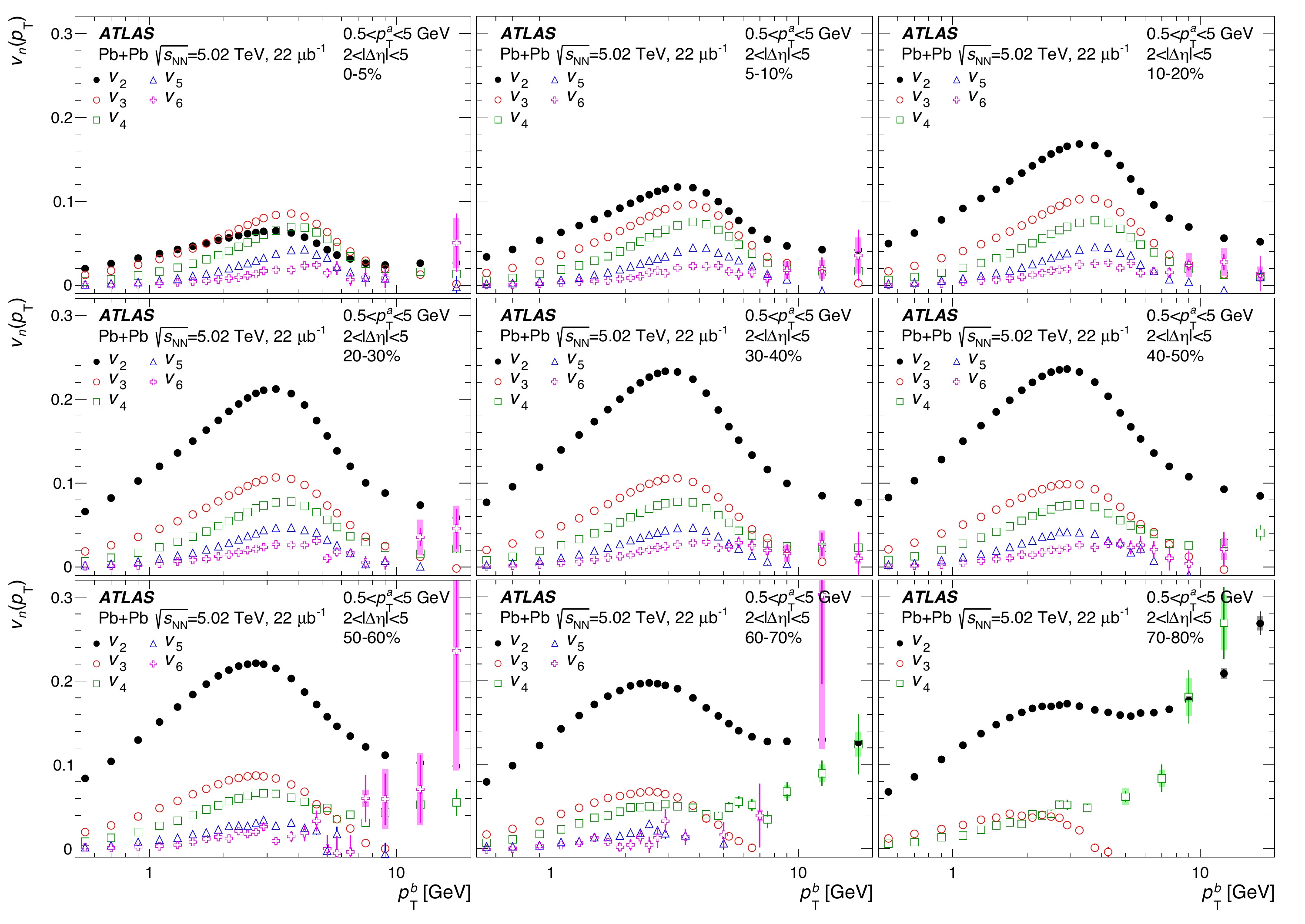}
\caption{
The \vn values obtained with the \tpc method as a function of \ptb for 0.5~<~\pta<~5~\GeV.
Each panel represents a different centrality interval.\
The vertical error bars indicate statistical uncertainties.\
The shaded bands indicate systematic uncertainties.
}
\label{fig:vn2PC_pt}
\end{figure}

\subsection{The \pt\ dependence of \vn}
Figures~\ref{fig:vnSP_pt} and~\ref{fig:vn2PC_pt} show the \vn obtained from the 
   SP and \tpc methods, respectively, as a function of \pT\ for several centrality 
   intervals.\
For the SP method the  \vtwo--\vfive harmonics are also shown for the 
   \cent{0--0.1} and \cent{0--1} ultra-central collisions.\
The SP results are integrated over the pseudorapidity $|\eta|<2.5$ and  the \tpc results are obtained with 0.5~<~\pta<~5~\GeV\ and for 2~<~|\deta|~<~5.\
The \vn values show a similar \pt\ dependence across all centralities: a nearly linear
   rise to about 2~\GeV, followed by a gradual increase to reach a maximum around 
   2--4~\GeV\ and a gradual fall at higher \pT.\
However, significant \vn values persist at high \pt ($\sim$20~\GeV).\
The \vtwo is positive even at the highest measured \pT of 60 \GeV\ 
  (Figure~\ref{fig:vnSP_pt}).\
This indicates the parton energy loss in the created medium~\cite{Sirunyan:2017pan}.\
Such elliptic flow is expected
due to path-length dependence of the energy loss of
high-\pt\ partons traversing the hot and dense medium.\
In peripheral events, at the highest \pt, the \tpc and SP \vtwo values again 
   show an increasing trend due to the increasing influence of 
   the away-side jet.\
The increased \vtwo in peripheral collisions at high-\pt\ 
   is accompanied by reduced values of \vthree and
   increased values of \vfour, which is characteristic of a large away-side 
   peak, as described in Section\ \ref{sec:2pc}.\
This is most clearly seen in the \cent{70--80} centrality interval.

The \vtwo varies significantly with centrality, reflecting a change in 
   the shape of the average initial collision geometry, from nearly 
   circular in ultra-central collisions to an almond shape in peripheral events.\
The higher harmonics do not show similar behaviour, as neither 
   higher-order eccentricities nor the fluctuations vary so 
   significantly with the centrality.\ 
The \vtwo is dominant at all centralities, except for the most central collisions
   interval where, at intermediate \pt, \vthree and \vfour become larger than \vtwo, 
   indicating that the dominant source  of observed flow comes from 
   the initial geometry fluctuations.\
This change in the \vn ordering is even more pronounced in the 1\% and 0.1\% 
   ultra-central collisions measured using the SP method, which shows that,  
   in the \pT region around the \vn peak, $\vthree > \vfour > \vtwo \approx \vfive$.\
The \vfour, similarly to \vtwo, exhibits an increase beyond  $\pT\sim10\ \GeV$,  
   which can be attributed to the presence of the events with dijets in the data.
In the SP measurement the \vseven results are also presented.\
The characteristics of \vseven are similar to the other 
  high-order harmonics, but the values are smaller 
  and significant, given the uncertainties, only in central and
  mid-central collisions and for the \pt\ range of 2--6~\GeV.\

\subsubsection{The Scalar Product and Event Plane methods comparison}

\begin{figure}[h!]
\centering
\includegraphics[width=0.8\textwidth]{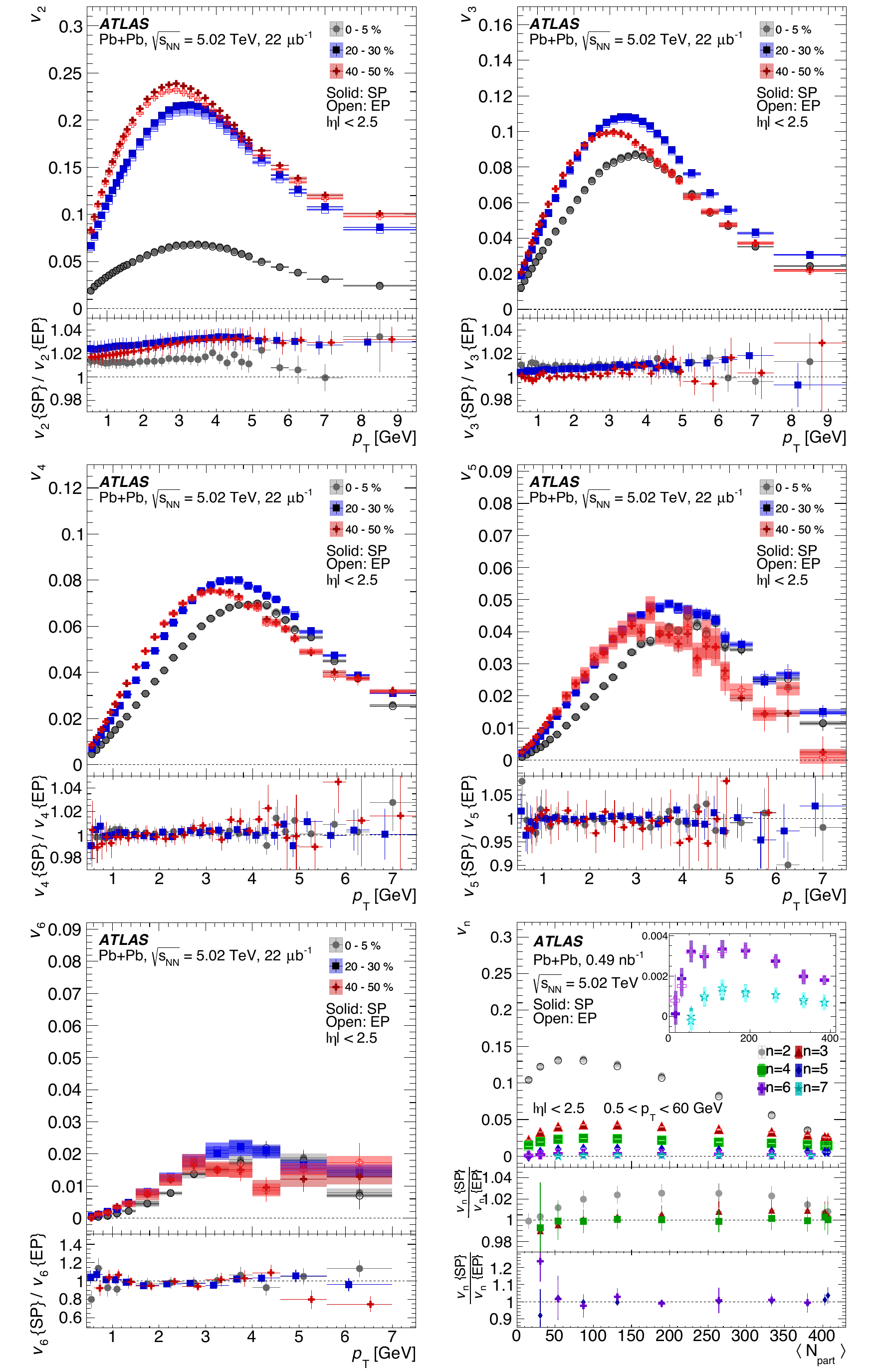}
\caption{
Comparison of the \vn\ obtained with EP and SP methods as a function of 
  \pT\ in three centrality bins: \cent{0--5}, \cent{20--30} 
  and \cent{40--50}.\
The right bottom panel shows the \vn\ as a function of $N_{\mathrm{part}}$, 
  integrated over 0.5~<~\pt~<~60~\GeV.\
   The correspondence of $N_{\mathrm{part}}$ to centrality intervals is provided in Table~\ref{tab:cent_def}.
In the inset the \vsix and \vseven integrated over 0.5~<~\pT~<~60~\GeV\ are 
  shown with adjusted scale.\
  For the $\vn(\pT)$ comparisons, the results are averaged 
  over the intervals indicated by horizontal error bars.\
The vertical error bars indicate the quadrature sum of statistical 
  and systematical uncertainties.
}
\label{fig:vnSP_EP_pt}
\end{figure}

Figure\,\ref{fig:vnSP_EP_pt} compares the \vn values measured with the EP and 
   SP methods as a function of \pT and $N_{\mathrm{part}}$ for the integrated 
   \pt\ range of $0.5<\pt<60$~\GeV.\ 
A small difference is seen between the \vtwo values measured with the two methods.\
The difference is largest in mid-central events: about 3\% in the \cent{20--30} 
   and \cent{40--50} centrality intervals, about 1\% in the \cent{0--5} most 
   central collisions and negligible in peripheral collisions.\ 
This difference is expected according to Ref.~\cite{Luzum:2012da} as the SP 
   method measures $\sqrt{\langle\vn^2\rangle}$ while the EP method measures values between 
   $\langle\vn\rangle$ and $\sqrt{\langle\vn^2\rangle}$, with the former value attained 
   in the limit of a small correction factor (the inverse of the denominator in
   Eq.~(\ref{eq:vnEP})) and approaching the latter when the correction factor is large.\ 
In the most central and peripheral events, where the correction is large for
   the second-order harmonic, the EP \vtwo values are closer to the SP ones, while
   for the mid-central events where the correction is small, the EP \vtwo 
   values are systematically lower than the SP \vtwo values.\
Higher-order EP and SP \vn harmonics are consistent with each other.

\subsubsection{The Scalar Product and two-particle correlation methods comparison}
\begin{figure}[h!]
\centering
\includegraphics[width=0.8\linewidth]{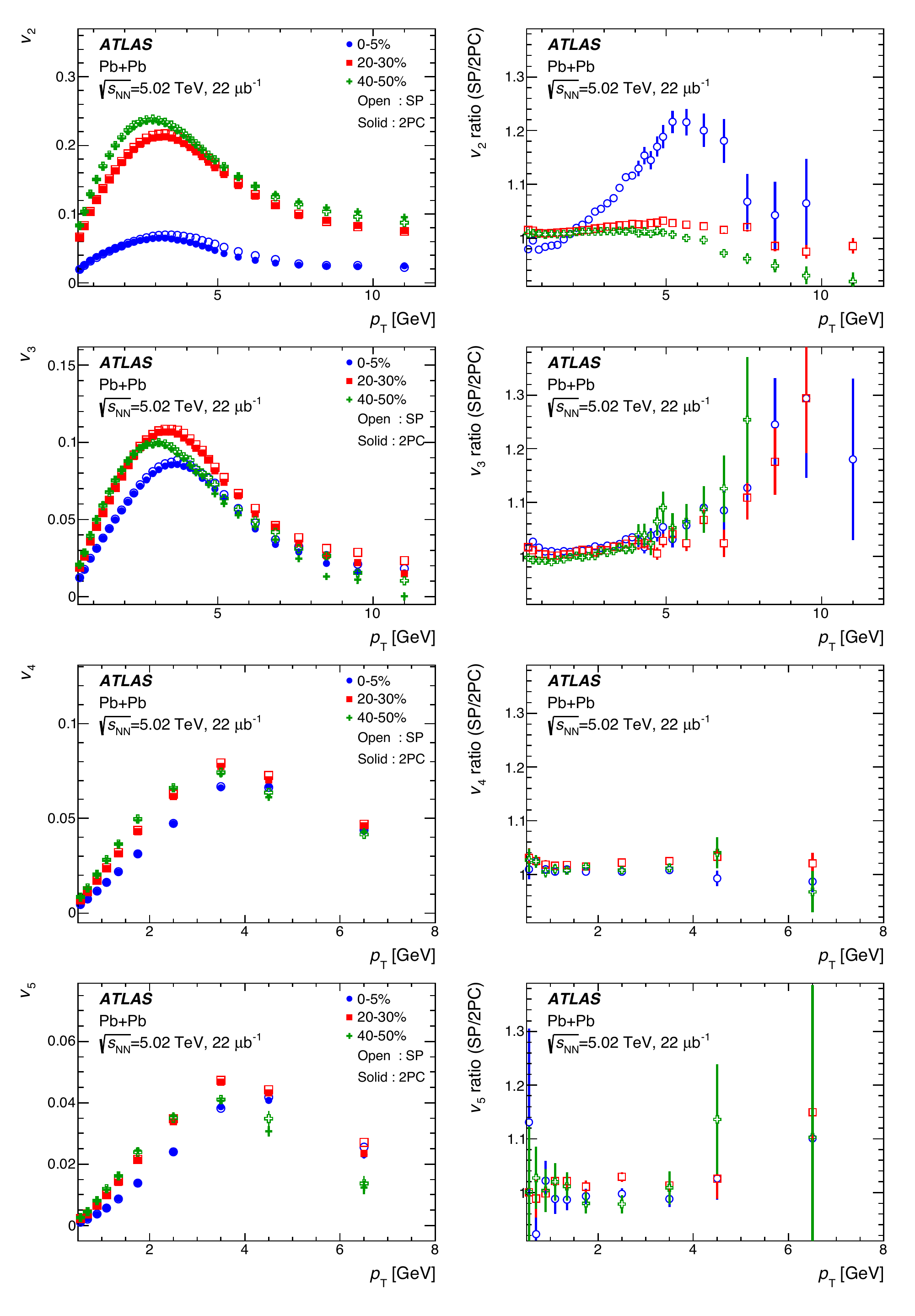}
\caption{
Comparison of the \vn\ obtained with \tpc and SP methods as a function of \pt.
Each panel shows the comparison for a different order harmonic.\
The comparisons are shown for three different centrality intervals: 
  \cent{0--5}, \cent{20--30} and \cent{40--50}.\
The vertical error bars indicate statistical uncertainties only.
}
\label{fig:vnSP_2PC_pt}
\end{figure}

A comparison of the SP and \tpc results is presented in Figure\,\ref{fig:vnSP_2PC_pt}.
In general,
results from the two methods are quite consistent.\
There is a significant difference in \vtwo from the two methods in the phase-space region $\pt<5$~\GeV,  0--5\% centrality.\
This difference decreases considerably for \cent{20--30} mid-central events, 
   where the \vtwo values match within 2--5\% up to $\pt\sim$10~\GeV.\
For \vthree--\vfive, where there are enough events for a clear comparison,
   the \vn values match within $\sim$4\% for $\pt<4$~\GeV\ for the three
   centrality intervals shown in Figure\,\ref{fig:vnSP_2PC_pt}.\
In principle, both the SP and \tpc methods measure $\sqrt{\langle\vn^2\rangle}$
   and the flow harmonics measured by the two methods should be identical.\
However, a breakdown of factorization (Eq.\,(\ref{eq:factorization})) results in systematic differences in the flow harmonics measurement.\
Such factorization breakdown has been observed to be significant for 
   \vtwo in central events~\cite{CMS-HIN-12-011}, 
   and in general for all \vn\ at higher \pt,  and is the leading
   source of disagreement between the 2PC and SP results.\
Furthermore, in the 2PC method the $\Delta\eta$ gap between the reference 
   and associated particles is chosen to be |\deta|~>~2, while in the 
   SP method, where the reference flow is measured in the FCal, the 
   minimum gap between the tracks and the FCal is 3.2 
   units in $\eta$.\ 
The presence of longitudinal-flow fluctuations, in which the event-plane angle 
   can change with $\eta$, can result in different \vn values depending on the $\eta$
   range where the reference flow is measured~\cite{Jia:2014vja,CMS-HIN-14-012}.\
This effect is also found to be larger in central events and relatively smaller
   in mid-central events~\cite{CMS-HIN-14-012}.\
These effects can further contribute to the observed difference between 
   the SP and \tpc \vn values.

\FloatBarrier
\subsubsection{Comparison to Pb+Pb results at \sqrtsnn~= 2.76~\TeV}
\begin{figure}[h!]
\includegraphics[width=0.99\linewidth]{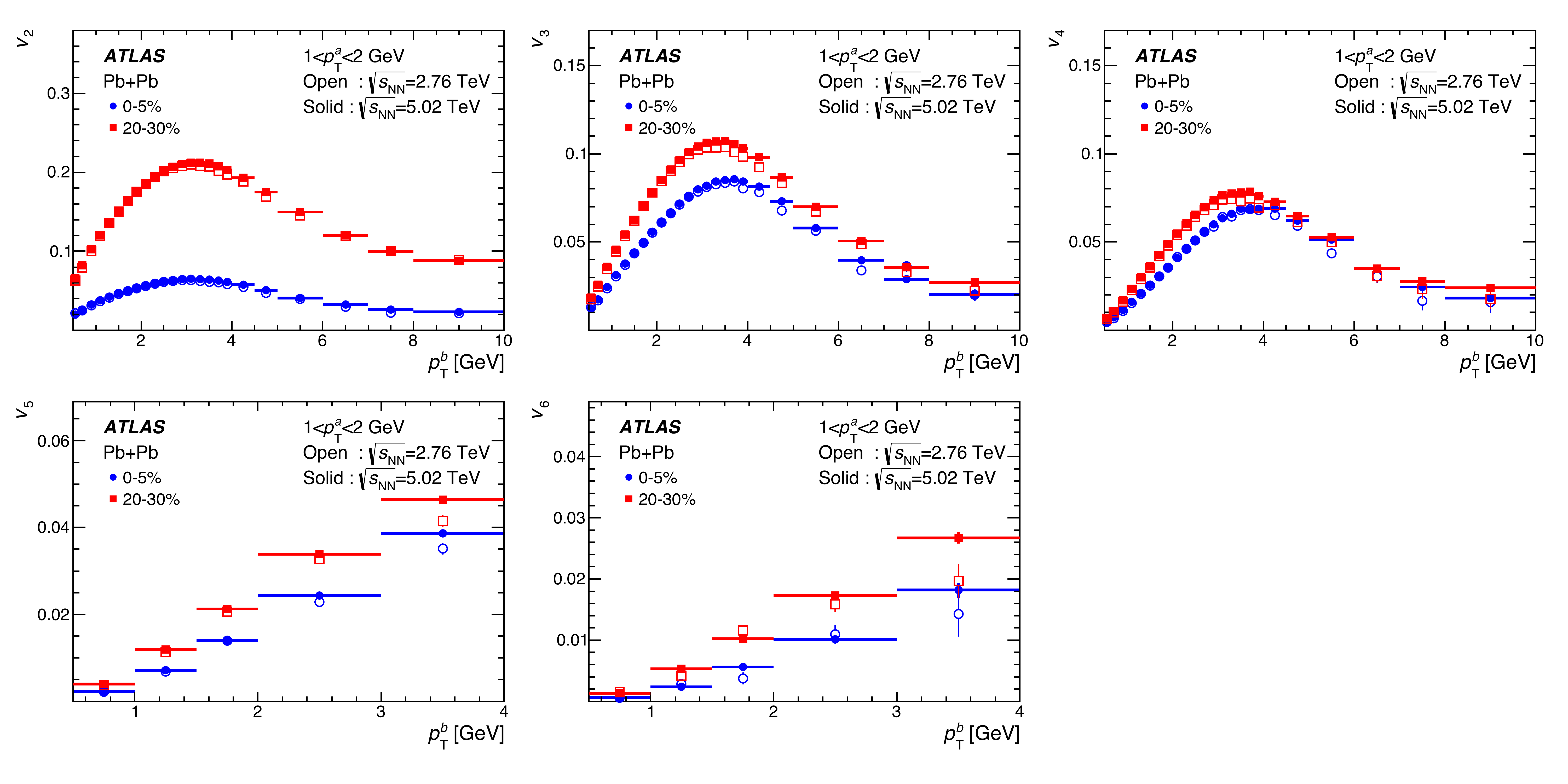}
\caption{
Comparisons of the \tpc \vn harmonics measured at \sqrtsnn~=~2.76~\TeV\ (Run\,1) and
   at \sqrtsnn~=~5.02~\TeV\ (Run\,2).\ 
The results are plotted as a function of \ptb for 1~<~\pta~<~2~\GeV\ for two centralities:
   \cent{0--5} and \cent{20--30}.\ 
Each panel corresponds to a different harmonic.\
Results are averaged over the intervals indicated by
  horizontal error bars.\
The vertical error bars indicate statistical uncertainties.
}
\label{fig:run1_run2_compare_2pc}
\end{figure}

\begin{figure}[h!]
\centering
\includegraphics[width=0.8\textwidth]{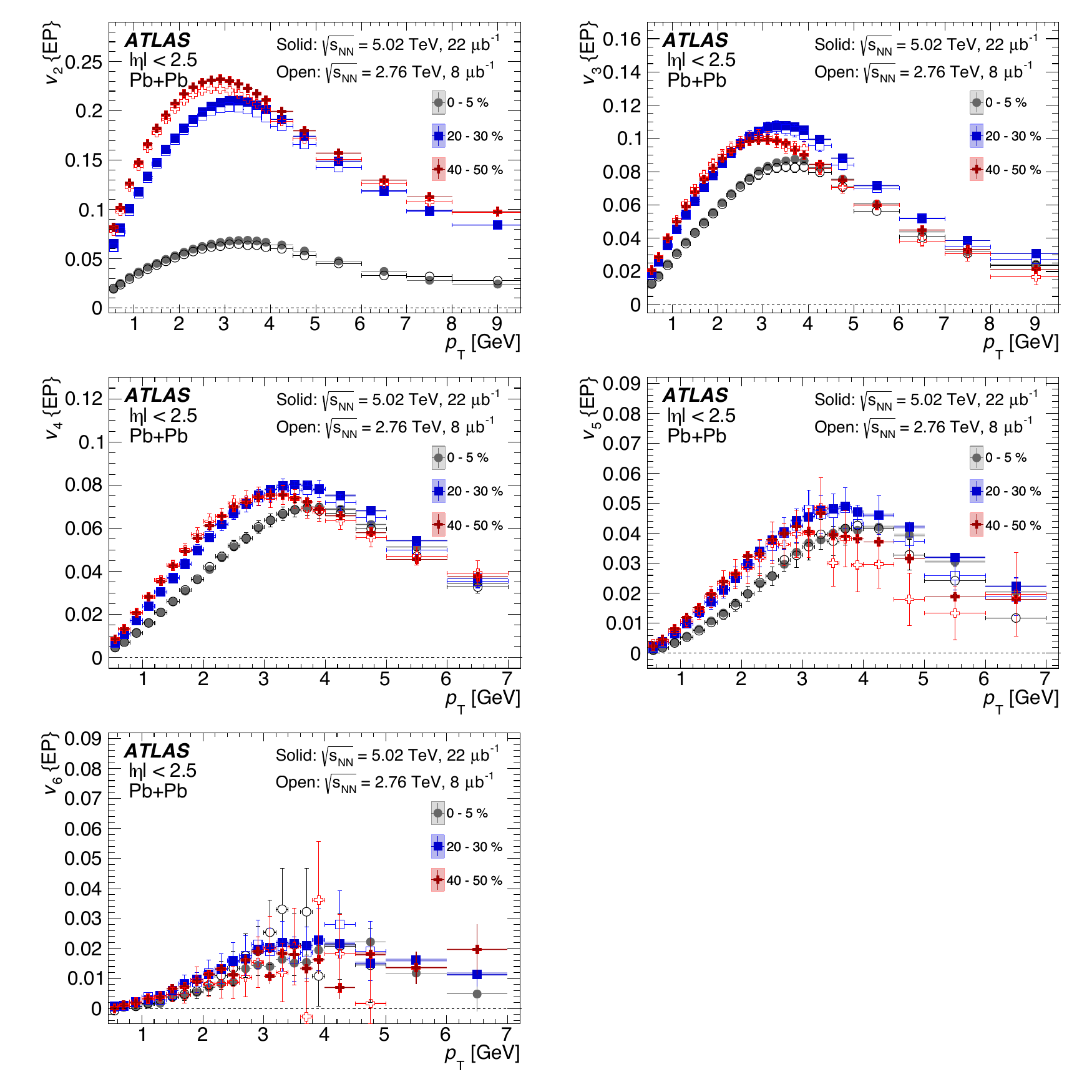}
\caption{
Comparison of the \vn\ obtained with EP method using Run\,1 and Run\,2 
  data as a function of \pT.\
The results are shown in three centrality intervals: \cent{0--5}, 
  \cent{20--30} and \cent{40--50}.\
Results are averaged over the intervals indicated by horizontal error bars.\
The vertical error bars indicate statistical uncertainties.\
The shaded areas indicate systematic uncertainties.
}
\label{fig:run1_run2_compare_ep}
\end{figure}

Figure\,\ref{fig:run1_run2_compare_2pc} shows a comparison of the \vn
   measured in the present analysis at $\sqrtsnn=5.02$~\TeV\ with the
   corresponding measurements at $\sqrtsnn=2.76$~\TeV\ for harmonics
   \vtwo to \vsix obtained using the 2PC method~\cite{HION-2011-01}.\
The comparisons are shown for two centralities: a central interval of \cent{0--5} and a 
   mid-central interval of \cent{20--30}.\
Figure~\ref{fig:run1_run2_compare_ep} shows a similar comparison of
  results obtained using the EP method for \cent{0--5}, \cent{20--30} and  \cent{40--50}
  centrality bins.\
The \vn at the two energies are quite similar and almost consistent throughout 
    within systematic and statistical uncertainties, even though the MC  
   non-closure correction was applied only in the $\sqrtsnn=5.02$~\TeV\ measurement.\
These results are consistent with the recent ALICE measurements 
   comparing the measurement of \vn
   at the two collision energies~\cite{Adam:2016izf}. 
 
\clearpage

\subsection{The $\eta$ dependence of \vn}

\begin{figure}[ht!]
\centering
\includegraphics[width=\textwidth]{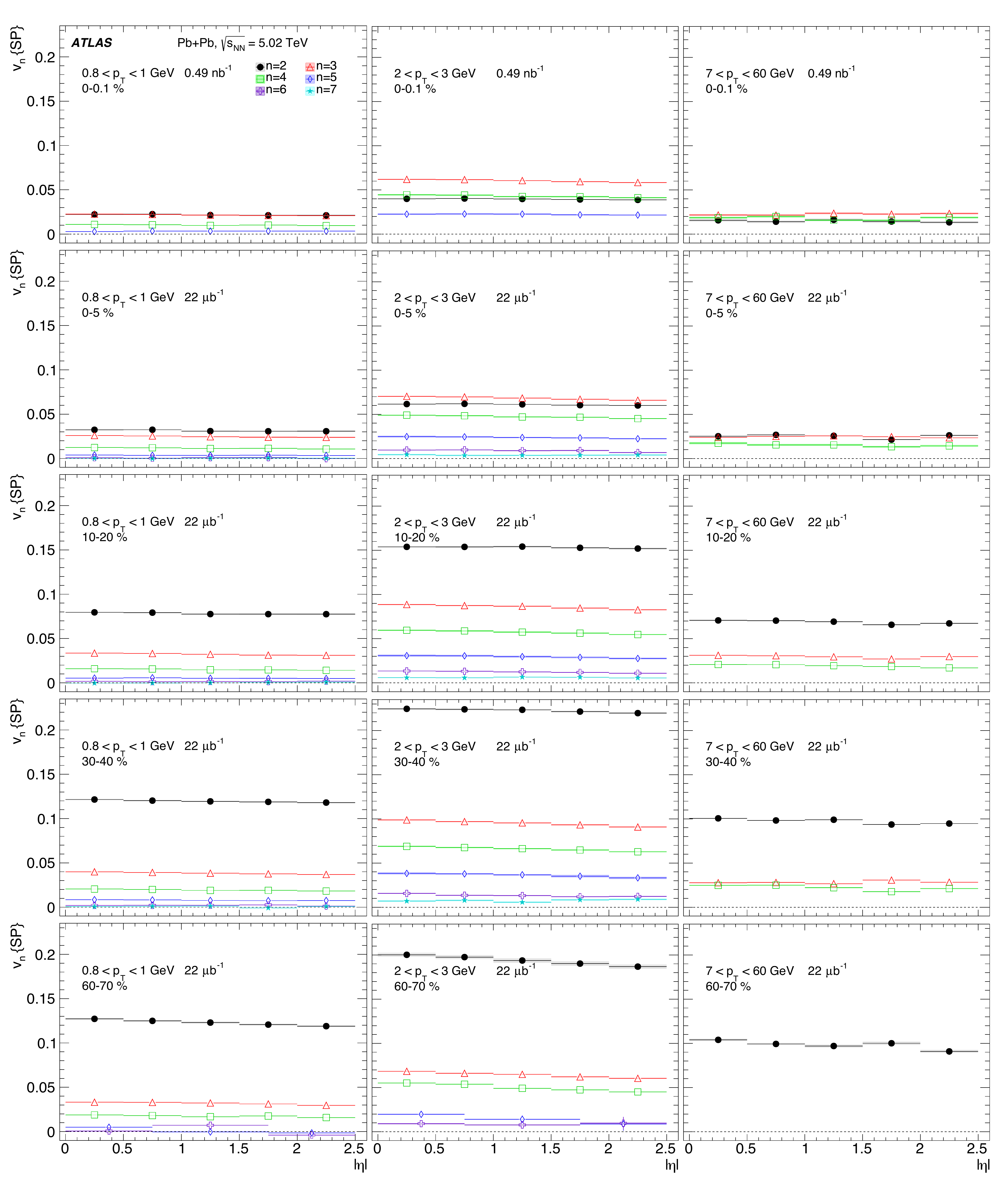}
\caption{
The \vn as a function of pseudorapidity obtained with the SP method, 
  for transverse momentum ranges: $0.8<\pT<1$~\GeV (left column), 
  $2<\pT<3$~\GeV (middle column) and $7 < \pT < 60$~\GeV (right column) 
  and for centrality intervals: \cent{0--0.1}(top row), \cent{0--5}, 
  \cent{10--20}, \cent{30--40} and \cent{60--70}(bottom row).\
The vertical error bars indicate statistical uncertainties.\
The shaded boxes indicate systematic uncertainties.
}
\label{fig:vnSP_eta}
\end{figure}

The pseudorapidity dependence of the \vtwo--\vseven obtained from the 
  SP method is shown in Figure\,\ref{fig:vnSP_eta} as a function of $|\eta|$.\
Benefiting from the symmetry of $\vn(\eta)$  with respect to $\eta=0$, 
  the \vn pseudorapidity dependence over the full range of pseudrapidity 
  was folded into the $\eta$ range 0--2.5.\ 
The $\eta$-dependence is shown for three ranges of transverse momenta 
  $0.8<\pT<1$~\GeV, $2<\pT<3$~\GeV\ and $7<\pT<60$~\GeV\ and for five 
  centrality intervals \cent{0--0.1}, \cent{0--5}, \cent{10--20}, 
  \cent{30-40} and \cent{60--70}.\
The strong dependence of flow harmonics on \pT and centrality shown across different panels (all vertical axes 
  in Figure\,\ref{fig:vnSP_eta} have the same range) is 
  discussed in the previous section.\ 
On the other hand, no strong pseudorapidity dependence of \vn harmonics 
  is observed.\ 
The \vtwo harmonic in central and mid-central collisions for $\pt<3$~\GeV\ 
  drops only by about 2--4\% over the pseudorapidity range $|\eta|$ = 0--2.5.\
For peripheral collisions and for high $\pt>7$~\GeV\ a larger decrease of 
  about 10\% is observed.\
The \vthree harmonic in central and mid-central collisions over the \pT\ range from 
  2 to 3~\GeV\  decreases by about 10\% with a larger drop of $\sim$15\% for 
  peripheral collisions.\
Similar pseudorapidity dependence is measured for the \vfour harmonic in 
  central and mid-central collisions over the \pT\ range from 2 to 3~\GeV\ 
  where it changes by about 10\%, but a larger drop of 25\% is observed in 
  peripheral collisions.\
In other cases, \vn harmonics are almost consistent with a uniform distribution 
  within the statistical and systematic uncertainties.\
As observed in the earlier measurement of \vn harmonics in 2.76~\TeV\ 
  \PbPb collisions~\cite{HION-2012-04}, such a weak $\eta$ dependence 
  of \vtwo may be partially attributed to a contribution of ``non-flow'' 
  short-range two-particle correlations.\

\subsection{Centrality dependence of \vn}
\begin{figure}[h!]
\centering
\includegraphics[width=0.9\textwidth]{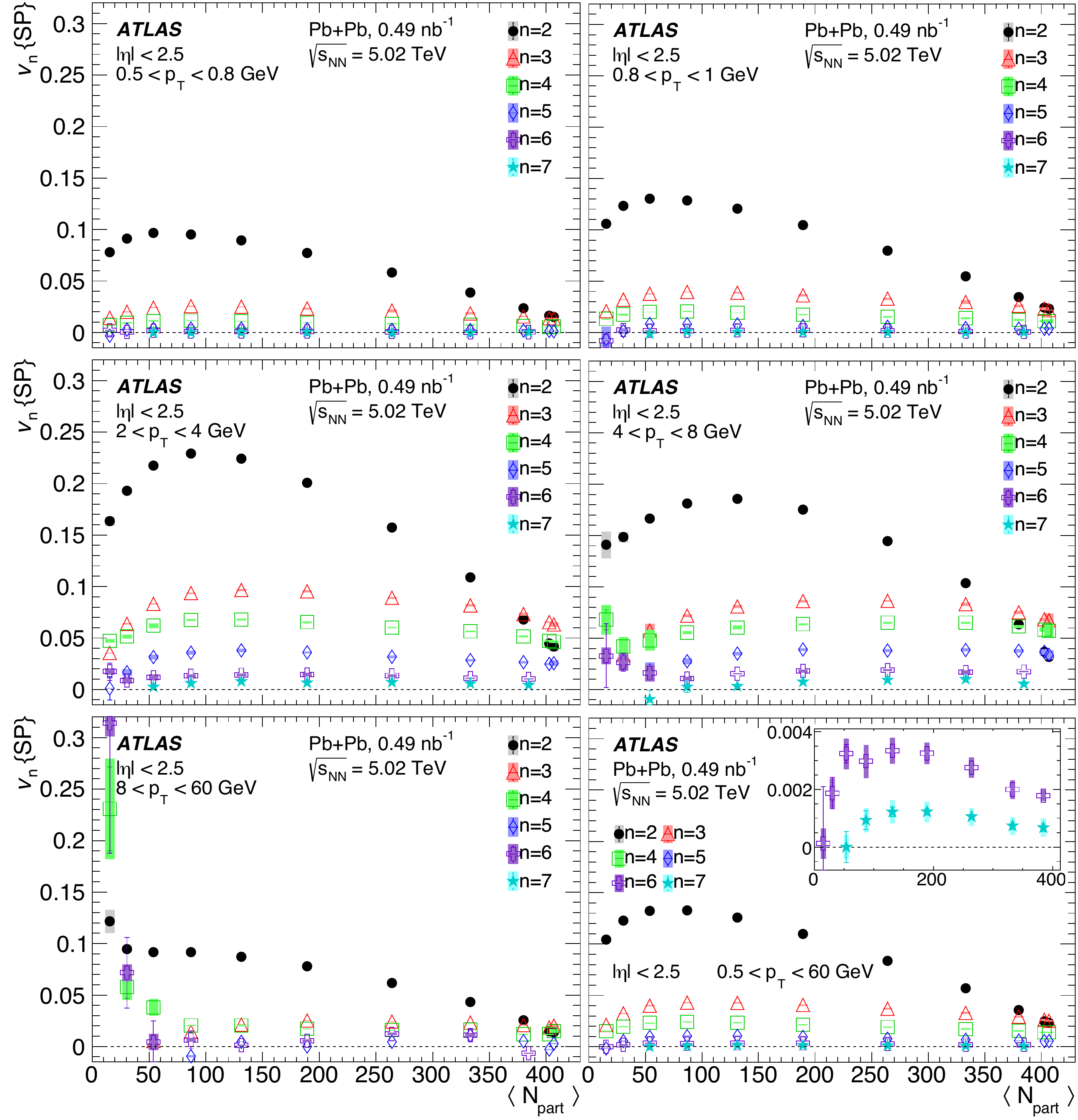}
\caption{
Integrated $v_{n}\{\mathrm{SP}\}$ vs. $N_{\mathrm{part}}$ for six \pT ranges shown in the panels 
  from lowest \pT range at the top left to the highest at the
  bottom middle, measured using the scalar-product method.\ 
In the inset in the bottom-right panel  the \vsix and \vseven integrated
over $0.5<\pT<60~\GeV$ are shown with adjusted scale.\
The correspondence of $N_{\mathrm{part}}$ to centrality intervals is provided in Table~\ref{tab:cent_def}.
Results are averaged over the intervals indicated by
  horizontal error bars.\
The vertical error bars indicate statistical uncertainties.\
The shaded areas indicate systematic uncertainties.
\label{fig:vnSP_centrality}}
\end{figure}

Figure~\ref{fig:vnSP_centrality} shows the $N_{\mathrm{part}}$ dependence 
   of \vn integrated over $|\eta|<2.5$ and for various ranges of 
   \pT using the SP method.\
   The elliptic flow is the dominant anisotropy, except at the largest $N_{\mathrm{part}}$ ($N_{\mathrm{part}} \gtrsim $~350), which corresponds to the 0-5\% most central collisions.\
For $\pT<8~\GeV$, a clear dependence on initial geometry can be observed as  
   \vtwo\ is highest in mid-central collisions, where this asymmetry
   is most significant.\ 
For $\pT>8$ \GeV, \vtwo is still the dominant harmonic, and it is 
   non-zero even in peripheral collisions as non-flow effects
   start to contribute in this region.\
A hierarchy $v_{n+1} < v_n$ is observed for $n=2$--$7$ for all 
   ranges of \pT\ and all centralities, except for the two bins \cent{0--0.1} 
   and \cent{0--1} of the ultra-central collisions at intermediate \pT,
   which are characterised by the flow harmonics ordering 
   $\vthree > \vfour > \vtwo  \approx \vfive$.

\FloatBarrier

\FloatBarrier

\subsection{$\vn(\pt)$ scaling\label{sec:vn_scaling}}

\begin{figure}[h!]
\centering
\includegraphics[width=0.9\linewidth]{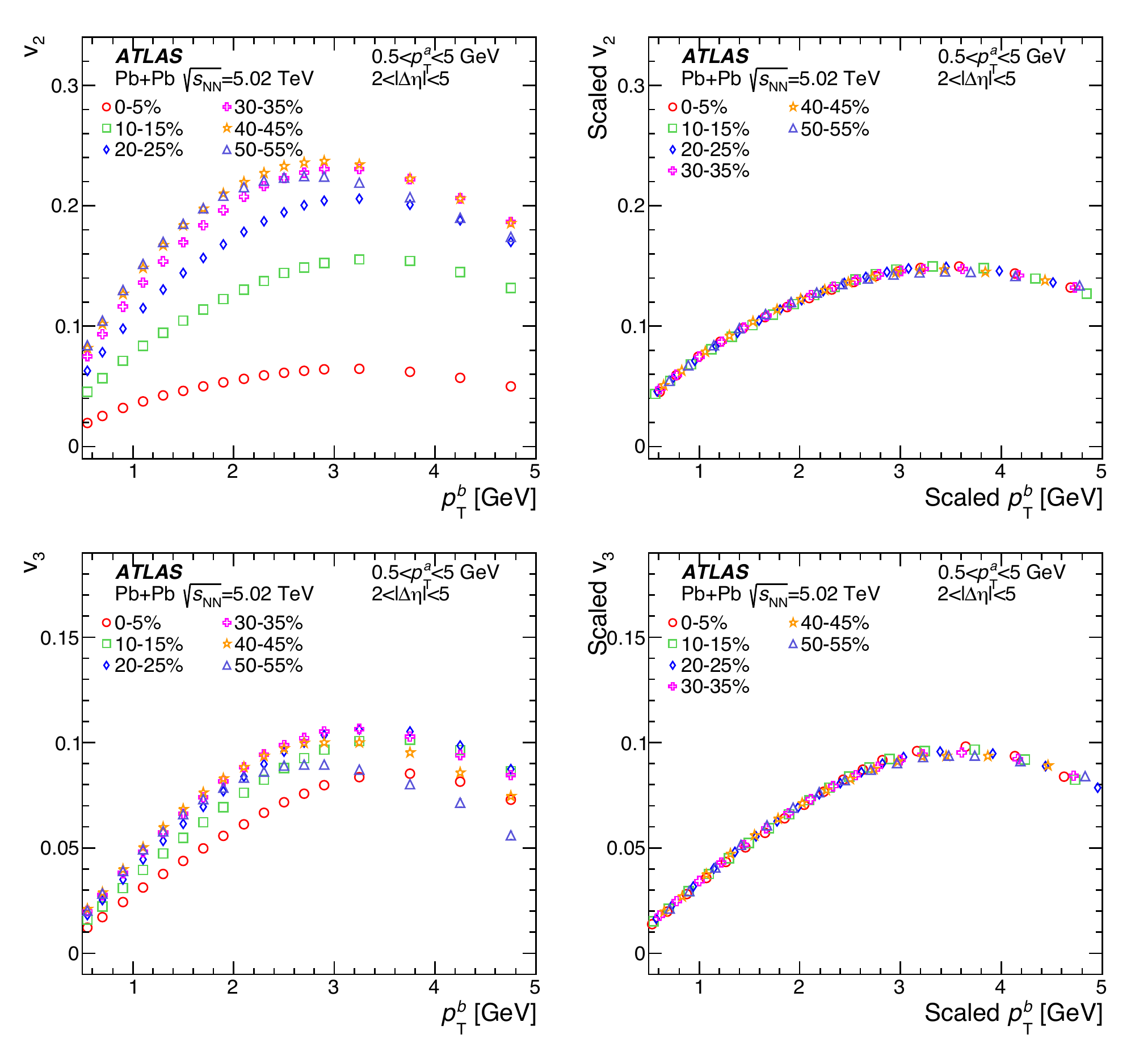}
\caption{
Left panels: the \vtwo(\pt) (top) and \vthree(\pt) (bottom) for
  different centrality intervals.\
Right panels: the corresponding scaled-$v_{n}(\pt)$.
The error bars indicate statistical and systematic 
  uncertainties added in quadrature and are typically 
  too small to be seen.
\label{fig:vn_scaled}}
\end{figure}

The left panels of Figure~\ref{fig:vn_scaled} compare the \pt\ dependence 
  of \vtwo and \vthree for different centrality intervals obtained from
  the 2PC method.\
Two distinct features are observed
  to change in the shape of the $\vn(\pt)$ (for the same $n$) between 
  the different centrality intervals: 
\begin{enumerate}
\item Change in the $v_{n}$-scale: The overall magnitude of the $\vn$
        changes from one centrality to another.\
      This effect is particularly large for $v_{2}$, and has to do 
        with the changing of the average collision geometry from one 
        centrality to another.
\item Change in the $\pt$ scale: The $\pt$ value at which the $\vn$ 
        reaches its maximum also changes systematically from one 
        centrality to another.
\end{enumerate}
In a recent ATLAS paper~\cite{HION-2013-04} it was observed that for a 
  given order $n$, the $\vn(\pt)$ in $p$+Pb and Pb+Pb  collisions have 
  a very similar $\pt$ dependence.\ 
In fact, after a scaling along the $\pt$ axis to account for the 
  difference in $\langle\pt\rangle$ between $p$+Pb and Pb+Pb 
  collisions, and an empirical scaling along the $\vn$ (i.e.\ $y$-axis) 
  to account for the difference in the collision geometry between the 
  two systems, the scaled $\vn(\pt)$ in $p$+Pb and Pb+Pb collisions were 
  found to be nearly identical.\ 
In this section, a check is done to see if such a scaling along the $\pt$ and 
  $\vn$ axes can yield universal shapes for the $\vn$ across the 
  different centrality classes.\
Accordingly, the $v_{n}(\pt)$ are scaled along the $x$- and $y$-axes to 
  match their shapes across the different centrality intervals.\
For the matching, the 0--60\% centrality interval is chosen
  as the reference, and the $\vn(\pt)$ for the individual 5\%-wide 
  centralities are scaled to match best the $\vn(\pt)$ in the 
  0--60\% centrality interval over the 0.5--5~\GeV\ $\pt$ range,
  with the scales along the $x$- and $y$-axes treated as fit parameters.\
The right panels of Figure~\ref{fig:vn_scaled} show the scaled-$v_{n}$, for $n=2$ and 3.\
Overall, the $\vn(\pt)$ shapes match well after the $\pt$ and $\vn$ scalings.

\begin{figure}[h!]
\centering
\includegraphics[width=1\linewidth]{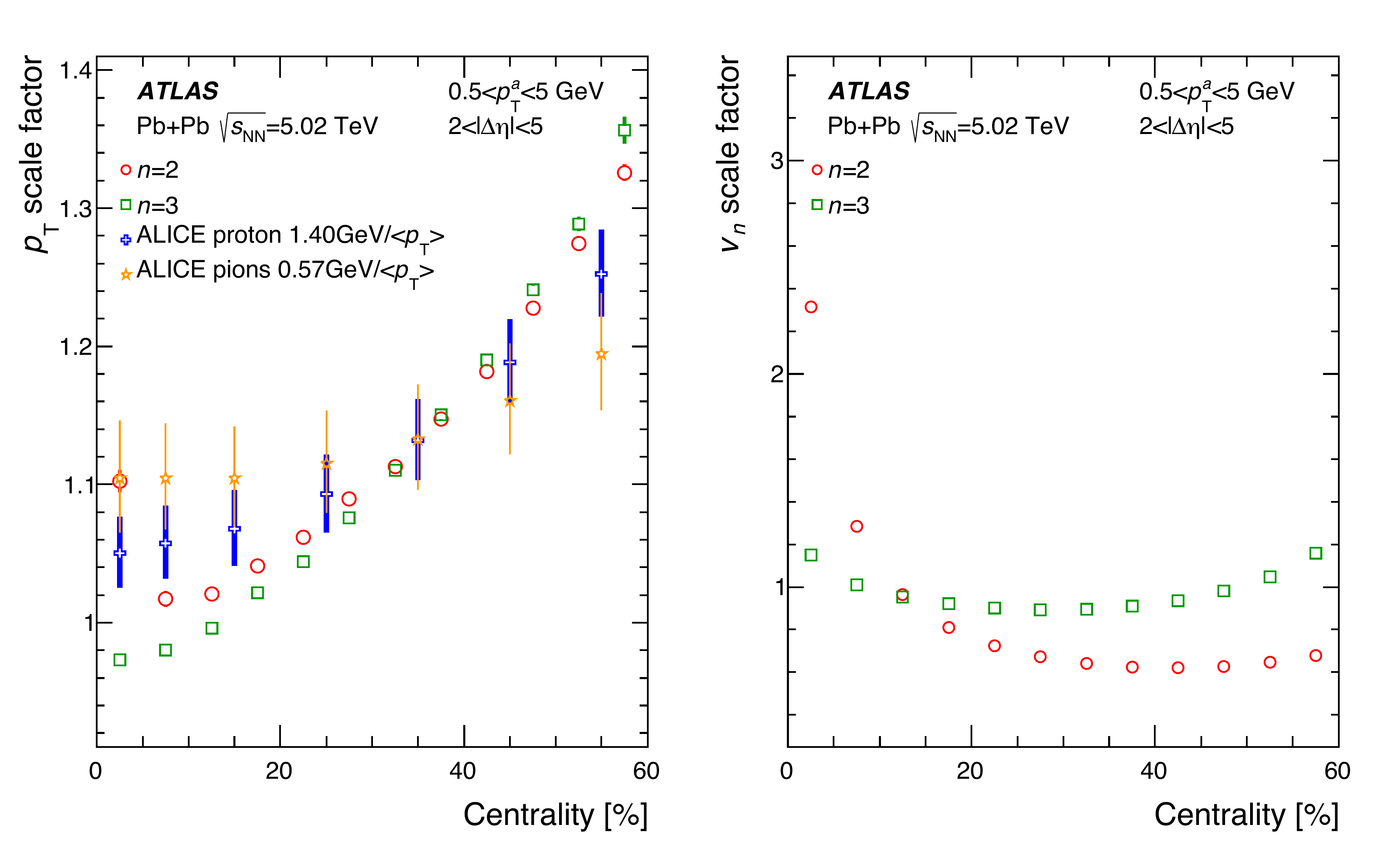}
\caption{
Left panel: the $x$-scale factors for the $v_{2}(\pt)$ and $v_{3}(\pt)$
  (see text) as a function of the collision centrality.\
Also shown for comparison are the ALICE 1/$\langle\pt\rangle$ for positively 
  charged protons and pions 
  (scaled by constant factors of 1.4~\GeV\ and 0.57~\GeV, respectively for plotting purposes). 
Right panel: the $y$-scale factors.\ 
The error bars on the scale factors and on the ALICE data indicate 
  systematic and statistical uncertainties added in quadrature. 
\label{fig:scale_factors}}
\end{figure}

Figure~\ref{fig:scale_factors} shows the $x$ (or $\pt$) and $y$-scale
  factors obtained for $v_{2}$ and $v_{3}$ as a function of centrality.\
It is interesting to note that the $\pt$-scale factors are quite
  comparable between $v_{2}$ and $v_{3}$ across most of the centrality
  ranges.\
However, in the 0--10\% most central events, some significant
  difference is observed between the two scale factors.\ 
This difference could be due to larger jet-bias and factorization-breaking effects in the $v_{2}$ as compared to $v_{3}$.\ 
The right panel of Figure~\ref{fig:scale_factors} shows the $y$-scale
  factors for $v_{2}$ and $v_{3}$ as well.\ 
Their centrality dependence is very different for the two harmonics.\ 
This is to be expected as the $y$-scale factors are mostly indicative 
  of the changing collision geometry, which becomes more and more 
  elliptic from central to mid-central events resulting in a large 
  increase in \vtwo, while \vthree, which is driven by 
  fluctuations, changes only gradually.

In order to check if the change in $\langle\pt\rangle$ between the 
  different centralities accounts for the change in the $x$-scale 
  of the $\vn$, the $1/\langle\pt\rangle$ for protons and pions, 
as  measured by the ALICE Collaboration~\cite{Abelev:2013vea}, are 
  also shown for comparison.\
While the centrality dependence of these $1/\langle\pt\rangle$ factors
  is qualitatively similar to the scale factors for the $\vn$,
  the relative variation in $1/\langle\pt\rangle$ is significantly 
  smaller than the variation in the $x$-scale of the \vn, indicating that there are 
  additional effects at play besides the change in the mean $\pt$.\
However, whatever the origin of these effects may be, they result in 
  nearly identical scaling factors for $v_{2}$ and $v_{3}$.

\section{Summary}
\label{sec:summary}
This \papertype presents the first ATLAS measurements of 
   azimuthal anisotropy of charged particles in \PbPb collisions at 5.02~\TeV\ collected during LHC running 
   in 2015.\
The measurements are based on the \PbPb sample of 0.49~$\mathrm{nb}^{-1}$ integrated luminosity and are performed with the two-particle correlation, scalar-product and event-plane methods.\
The azimuthal anisotropy is quantified by the flow harmonics $\vtwo$--$\vsix$ and  $\vtwo$--$\vseven$ for measurements based on the 2PC and SP/EP methods, respectively.\ 
The flow harmonics are obtained in wide transverse momentum 
   (0.5 $< \pT < $ 60~\GeV), pseudorapidity ($|\eta|< 2.5$) and 
   centrality \cent{0--80} ranges.\
 All harmonics show a similar trend in the \pT dependence; first increasing with \pT 
   up to a maximum around 3--4 GeV and then decreasing for 
   higher \pT.\
 Significant values of the second-order harmonic \vtwo persist
   up to 60~\GeV.\
The \vtwo results at high \pt provide a useful handle for the study of
    partonic energy loss in the dense medium, and so can improve our 
    understanding of the QGP properties.\
    The values of the flow harmonics  decrease strongly with increasing harmonic order for all centralities, except for 0--5\% central collisions where a different ordering is seen:
 $\vthree > \vfour > \vtwo \approx \vfive$
for  \pT above 1 \GeV, 
which is indicative of the presence of significant \vtwo fluctuations in these ultra-central collisions.\
The elliptic flow signal is strongly dependent on event centrality 
and is largest in mid-central collisions of \cent{30--50}.\
The higher-order harmonics show a weak centrality dependence, which is 
   consistent with an anisotropy associated with fluctuations in the 
   initial geometry.\
After scaling the \pT and magnitude of the differential elliptic and 
  triangular, \vthree,  flow harmonics at different centralities to 
  best match the reference \vtwo and \vthree for the \cent{0--60} 
  centrality interval over the \pT range 0.5~$<\pT<$~5~GeV, the shapes 
  of the rescaled harmonics agree well, indicating similarity among \pT-shapes of flow harmonics evolving from different 
  initial conditions.\
The \vn coefficients are shown to exhibit a weak $\eta$-dependence, irrespective of the harmonic order, centrality and \pT.
A weaker pseudorapidity dependence of $\vn(\eta)$ in more central collisions is suggested by data.
The results obtained using the EP and SP methods are consistent for
   harmonics of order $n\geq3$.\
A small, systematic difference is observed for \vtwo, where the values 
   obtained from the SP method are up to 3\% larger than the values 
   obtained using the EP method.\
The \tpc and SP methods give values for \vn that are quite consistent 
   up to $\sim$10~\GeV.\
However, in the most central events the SP method gives 
   systematically larger values for \vtwo for \pt> 2~\GeV.\
Comparisons with measurements in \PbPb collisions at 
   $\sqrtsnn=2.76~\TeV$\ show that the \pT dependence of the \vn shows no change
   from $\sqrtsnn=2.76~\TeV$\ to $\sqrtsnn=5.02~\TeV$. 

The set of results on flow harmonics presented in this paper provides a tool for studies of the underlying mechanism leading to the large azimuthal anisotropy observed in the \PbPb collision system,
and can be used to constrain the theoretical modelling of the initial state of the QGP, its subsequent hydrodynamic evolution as well as partonic energy loss in the dense and hot medium.

\clearpage


\section*{Acknowledgements}

We thank CERN for the very successful operation of the LHC, as well as the
support staff from our institutions without whom ATLAS could not be
operated efficiently.

We acknowledge the support of ANPCyT, Argentina; YerPhI, Armenia; ARC, Australia; BMWFW and FWF, Austria; ANAS, Azerbaijan; SSTC, Belarus; CNPq and FAPESP, Brazil; NSERC, NRC and CFI, Canada; CERN; CONICYT, Chile; CAS, MOST and NSFC, China; COLCIENCIAS, Colombia; MSMT CR, MPO CR and VSC CR, Czech Republic; DNRF and DNSRC, Denmark; IN2P3-CNRS, CEA-DRF/IRFU, France; SRNSFG, Georgia; BMBF, HGF, and MPG, Germany; GSRT, Greece; RGC, Hong Kong SAR, China; ISF and Benoziyo Center, Israel; INFN, Italy; MEXT and JSPS, Japan; CNRST, Morocco; NWO, Netherlands; RCN, Norway; MNiSW and NCN, Poland; FCT, Portugal; MNE/IFA, Romania; MES of Russia and NRC KI, Russian Federation; JINR; MESTD, Serbia; MSSR, Slovakia; ARRS and MIZ\v{S}, Slovenia; DST/NRF, South Africa; MINECO, Spain; SRC and Wallenberg Foundation, Sweden; SERI, SNSF and Cantons of Bern and Geneva, Switzerland; MOST, Taiwan; TAEK, Turkey; STFC, United Kingdom; DOE and NSF, United States of America. In addition, individual groups and members have received support from BCKDF, CANARIE, CRC and Compute Canada, Canada; COST, ERC, ERDF, Horizon 2020, and Marie Sk{\l}odowska-Curie Actions, European Union; Investissements d' Avenir Labex and Idex, ANR, France; DFG and AvH Foundation, Germany; Herakleitos, Thales and Aristeia programmes co-financed by EU-ESF and the Greek NSRF, Greece; BSF-NSF and GIF, Israel; CERCA Programme Generalitat de Catalunya, Spain; The Royal Society and Leverhulme Trust, United Kingdom. 

The crucial computing support from all WLCG partners is acknowledged gratefully, in particular from CERN, the ATLAS Tier-1 facilities at TRIUMF (Canada), NDGF (Denmark, Norway, Sweden), CC-IN2P3 (France), KIT/GridKA (Germany), INFN-CNAF (Italy), NL-T1 (Netherlands), PIC (Spain), ASGC (Taiwan), RAL (UK) and BNL (USA), the Tier-2 facilities worldwide and large non-WLCG resource providers. Major contributors of computing resources are listed in Ref.~\cite{ATL-GEN-PUB-2016-002}.

\printbibliography
\clearpage
 
\begin{flushleft}
{\Large The ATLAS Collaboration}

\bigskip

M.~Aaboud$^\textrm{\scriptsize 34d}$,    
G.~Aad$^\textrm{\scriptsize 99}$,    
B.~Abbott$^\textrm{\scriptsize 124}$,    
O.~Abdinov$^\textrm{\scriptsize 13,*}$,    
B.~Abeloos$^\textrm{\scriptsize 128}$,    
D.K.~Abhayasinghe$^\textrm{\scriptsize 91}$,    
S.H.~Abidi$^\textrm{\scriptsize 164}$,    
O.S.~AbouZeid$^\textrm{\scriptsize 39}$,    
N.L.~Abraham$^\textrm{\scriptsize 153}$,    
H.~Abramowicz$^\textrm{\scriptsize 158}$
% ...
\bigskip
\\

% ...
$^{13}$Institute of Physics, Azerbaijan Academy of Sciences, Baku; Azerbaijan.\\
% ...
$^{34}$$^{(a)}$Facult\'e des Sciences Ain Chock, R\'eseau Universitaire de Physique des Hautes Energies - Universit\'e Hassan II, Casablanca;$^{(b)}$Centre National de l'Energie des Sciences Techniques Nucleaires (CNESTEN), Rabat;$^{(c)}$Facult\'e des Sciences Semlalia, Universit\'e Cadi Ayyad, LPHEA-Marrakech;$^{(d)}$Facult\'e des Sciences, Universit\'e Mohamed Premier and LPTPM, Oujda;$^{(e)}$Facult\'e des sciences, Universit\'e Mohammed V, Rabat; Morocco.\\
% ...
$^{39}$Niels Bohr Institute, University of Copenhagen, Copenhagen; Denmark.\\
% ...
$^{91}$Department of Physics, Royal Holloway University of London, Egham; United Kingdom.\\
% ...
$^{99}$CPPM, Aix-Marseille Universit\'e, CNRS/IN2P3, Marseille; France.\\
% ...
$^{124}$Homer L. Dodge Department of Physics and Astronomy, University of Oklahoma, Norman OK; United States of America.\\
% ...
$^{128}$LAL, Universit\'e Paris-Sud, CNRS/IN2P3, Universit\'e Paris-Saclay, Orsay; France.\\
% ...
$^{153}$Department of Physics and Astronomy, University of Sussex, Brighton; United Kingdom.\\
% ...
$^{158}$Raymond and Beverly Sackler School of Physics and Astronomy, Tel Aviv University, Tel Aviv; Israel.\\
% ...
$^{164}$Department of Physics, University of Toronto, Toronto ON; Canada.\\
% ...
$^{a}$ Also at Borough of Manhattan Community College, City University of New York, NY; United States of America.\\
$^{b}$ Also at Centre for High Performance Computing, CSIR Campus, Rosebank, Cape Town; South Africa.\\
$^{c}$ Also at CERN, Geneva; Switzerland.\\
$^{d}$ Also at CPPM, Aix-Marseille Universit\'e, CNRS/IN2P3, Marseille; France.\\
$^{e}$ Also at D\'epartement de Physique Nucl\'eaire et Corpusculaire, Universit\'e de Gen\`eve, Gen\`eve; Switzerland.\\
% ...
$^{*}$ Deceased

\end{flushleft}


\end{document}